\documentclass[twocolumn,pra,superscriptaddress,showpacs]{revtex4}
\usepackage{epsfig,color}
\usepackage{times}
\usepackage{amsmath}
\usepackage{graphics}

\newcommand{\Li}{{}^{6}\textrm{Li}}
\newcommand{\K}{{}^{40}\textrm{K}}

\begin{document}

\title{Dimer, trimer and FFLO liquids in mass- and spin-imbalanced trapped binary mixtures in one dimension}

\author{M. Dalmonte}
\email[Corresponding author: ]{marcello.dalmonte@uibk.ac.at}
\affiliation{Institute for Quantum Optics and Quantum Information of the Austrian Academy of Sciences, University of Innsbruck, A-6020 Innsbruck, Austria}
\affiliation{Dipartimento di Fisica dell'Universit\`a di Bologna and INFN, via Irnerio 46, 40127 Bologna, Italy}

\author{K. Dieckmann}
\affiliation{Centre for Quantum Technologies, National University of Singapore, 3 Science Drive 2, 117543 Singapore}

\author{T. Roscilde}
\affiliation{Laboratoire de Physique, CNRS UMR 5672, Ecole Normale Superieure de Lyon, Universite de Lyon, 46 Allee d’Italie, Lyon, F-69364, France}

\author{C. Hartl}
\affiliation{Department of Physics and Arnold Sommerfeld Center for Theoretical Physics, Ludwig-Maximilians-Universit\"at M\"unchen, D-80333 M\"unchen, Germany}

\author{A.~E.~Feiguin}
\affiliation{University of Wyoming, Laramie, WY 82071, USA}

\author{U. Schollw\"ock}
\affiliation{Department of Physics and Arnold Sommerfeld Center for Theoretical Physics, Ludwig-Maximilians-Universit\"at M\"unchen, D-80333 M\"unchen, Germany}
\affiliation{Kavli Institute for Theoretical Physics, Kohn Hall, University of California, Santa Barbara, California 93106, USA}

\author{F. Heidrich-Meisner}
\affiliation{Department of Physics and Arnold Sommerfeld Center for Theoretical Physics, Ludwig-Maximilians-Universit\"at M\"unchen, D-80333 M\"unchen, Germany}
\affiliation{Kavli Institute for Theoretical Physics, Kohn Hall, University of California, Santa Barbara, California 93106, USA}

\date{\today}

\begin{abstract}
We present a systematic investigation of attractive binary mixtures in presence
of both spin- and mass-imbalance in one dimensional setups described by the Hubbard model. After discussing
typical cold atomic experimental realizations and the relation between microscopic and effective
parameters, we study several many-body features of trapped Fermi-Fermi
and Bose-Bose mixtures such as density profiles, momentum distributions and 
correlation functions by means of numerical density-matrix-renormalization-group
and Quantum Monte Carlo simulations. In particular, we focus on the stability of Fulde-Ferrell-Larkin-Ovchinnikov, 
dimer and trimer fluids in inhomogeneous situations, as typically realized in cold
gas experiments due to the harmonic confinement. We finally consider possible experimental signatures of these 
phases both in the presence of a finite polarization and of a finite temperature.
\end{abstract}

\pacs{71.10.Pm, 05.30.Jp, 37.10.Jk}

\maketitle


\section{Introduction}
\label{sec:intro}

The study of superfluidity of either bosons or fermions has been a central topic in the field of ultracold atomic gases, starting from seminal
experimental studies on the Bose-Einstein condensation of bosons \cite{anderson95,davis95,bradley95} and continuing with investigations of the BCS-BEC crossover \cite{bartenstein04,bourdel04,regal04,zwierlein04}. 
More recent experiments with two-component Fermi gases 
have  addressed the case of 
a finite population imbalanced, both in three spatial dimensions 
\cite{zwierlein06,zwierlein06a,partridge06,partridge06a} and in one
dimension \cite{liao10}.
Among the  goals of  these experiments is the search for the transition
from a fully paired superfluid phase to the normal state and, in particular,
for competing pairing states that survive a finite polarization. These include 
the Sarma state \cite{sarma63} or the Fulde-Ferrell-Larkin-Ovchinnikov (FFLO) state \cite{fulde64,larkin64}.

With the advances optical lattice engineering \cite{bloch08}, it has also become possible
to investigate fermionic pairing states within the framework of the so-called
 Hubbard model. For repulsive onsite interactions 
the hope is to reach sufficiently low temperatures to search for pairing states away from half filling \cite{kohl05,joerdens08,schneider08},
whereas in the case of attractive interactions, there is a natural tendency to pair formation. 
Besides experiments with homonuclear mixtures, where the pseudo-spin degree of freedom arises from 
preparing atoms in different hyper-fine states, there is also the possibility of working with 
heteronuclear mixtures such as, for instance, the ${}^{40}$K-${}^6$Li system \cite{taglieber08,wille08,voigt09,trenkwalder11}. In that case, 
one deals both finite mass imbalance and population imbalance.
On the other hand, bosonic gases in optical lattices have provided the first example
of a quantum phase transition from a Mott-insulator to a superfluid state in the context of ultra cold atoms~\cite{greiner02},
which was subsequently observed in one dimensional setups \cite{Stoeferle2004,haller2010} as well. In analogy with the fermionic case,
both homonuclear and heteronuclear bosonic mixtures have been realized \cite{catani2008,mckay2010,gadway2011}; from a 
theoretical viewpoint, the two-species Bose-Hubbard model encompasses a remarkably rich physics,
ranging from super-counterflow and antiferromagnetic phases in case of interspecies
repulsion \cite{kuklov2003,altman2003,guglielmino2011} to pair superfluidity and density-wave instabilities in the attractive regime \cite{mathey2007,
hu2009,roscilde2012}.

From a theoretical point of view,  
it is of advantage to consider the one-dimensional case, for which 
both powerful analytical and numerical methods are available that can provide us with practically
exact answers in many regimes. In ultra-cold atomic gases, it is also perfectly possible
to realize one-dimensional systems experimentally, both for bosons \cite{Stoeferle2004} and fermions \cite{liao10}, adding
strong motivation to study this case. Moreover one-dimensional systems have been demonstrated to harbor unconventional 
superfluid phases which are the central topic of this work.
In our work, we are interested in three types of superfluid states of two-component systems and their existence in the
attractive, asymmetric Hubbard model: (i) the more conventional fully paired phase, which can be considered  as a fluid of Cooper pairs, (ii)
superfluids of larger composite objects such as trimers, and (iii) 
the FFLO state in which a superfluid of pairs with finite center-of-mass momentum appears.
The first two cases, i.e., the equal density (or dimer fluid) and the trimer fluid,   can be realized in both Fermi-Fermi 
mixtures \cite{orso10, burovski09} and Bose-Bose mixtures in one-dimension
\cite{mathey2007} while the FFLO state exists only in the case of two-component Fermi mixtures (see \cite{feiguin11} and references therein).
In our study we use numerically exact methods -- density matrix renormalization group (DMRG)\cite{white92b, schollwoeck05,schollwoeck11} and stochastic series expansion Quantum Monte Carlo \cite{syljuasen02}--
to probe the stability of the above cited phases at a finite population and mass imbalance in the presence of
a harmonic trapping potential. Our study complements previous works that mostly focussed on the homogeneous case \cite{batrouni09,wang09,burovski09,orso10,roux11,roscilde2012,hu2009}. Moreover, we incorporate aspects that are typical of many experiments,
namely the possibility
of tuning the ratio of the effective masses through the depth of the optical lattice; the fact that in a mass imbalanced
system the two components usually experience a different trapping potential; 
the difficulty in controlling the population of the two components to an arbitrary degree of precision, resulting typically in a non-zero global polarization.
Finally, we also study the effect of a small but finite temperature on the shell structure and coherence properties of trapped mass-imbalanced two-component gases.

Here we summarize our main results. In the case of fermions and for the bulk system it is well-known that the fully paired phase is the ground state of a two-component Fermi gas
away from half filling~\cite{essler-book}. In the presence of a harmonic trap and at a finite population imbalance,  
the fully paired phase can only survive in the wings of the particle cloud at small polarizations \cite{orso07,hu07,casula08,hm10a}, as recently observed in Ref.~\onlinecite{liao10}. Here we show
 that adding mass imbalance to the system stabilizes the fully paired phase in the trap:
if the light particles are the majority species, the fully paired phase occupies the center of the trapped system, 
in agreement with the grand-canonical phase diagram of the asymmetric Hubbard model at a finite spin polarization \cite{wang09,orso10,roux11}.
We further show that a small temperature does not destroy the equal density phase. The main conclusion is therefore
that it is not necessary to enforce the condition of a perfectly balanced gas to observe and study properties of a
dimer fluid. This result applies to both fermions and hard-core bosons.

A particular feature of mass-imbalanced systems in one dimension is that they allow for the formation of bound states
of more than two components, which in the two-species Hubbard model with equal masses is forbidden. This can happen
for both fermions \cite{burovski09,orso10,roux11} and bosons \cite{burovski09}. The existence of these states
requires commensurate ratios of the densities of heavy particles ($n_{\downarrow}$) versus light ones ($n_{\uparrow}$). 
For instance, a trimer fluid can only be stable
 if $2n_{\uparrow}=n_{\downarrow}$ \cite{burovski09,orso10}, where $n_{\sigma}=N_{\sigma}/L$ is the density of the pseudo-spin $\sigma=\uparrow,\downarrow$ component in a system of length $L$ with $N_{\sigma}$ particles of species $\sigma$.  For the homogeneous system, this sets the global polarization to 
\begin{equation}\label{eq:comm}
P=(N_{\uparrow}-N_{\downarrow})/N= -1/3
\end{equation}
($N=N_{\uparrow}+N_{\downarrow}$)
and it is in fact possible to stabilize this phase in a harmonic trap by enforcing this condition globally as was shown in Ref.~\onlinecite{roux11}
at sufficiently low densities. Here we investigate the fate of the trimer fluid
upon deviating from the global polarization of $P=-1/3$, both for the case of fermions and bosons, finding that the trimer fluid rapidly gets pushed into
the wings of the gas and subsequently disappears even for very small deviations from Eq.~\ref{eq:comm}. Therefore, one needs to fine-tune the global polarization to precisely $P=-1/3$ to 
ensure that a large fraction of particles participates in this state.
A difference in the trapping potential for the two components, however, has little effect on the trimer phase. For the case of bosons,
we demonstrate that the formation of the trimer fluid leaves clear fingerprints in the momentum distribution
function, a quantity that is easily accessible in experiments.

Finally, concerning the FFLO state, we show that it survives the addition
of the harmonic trap and the mass imbalance. Similar to the case of a system with mass imbalance only \cite{orso07,hu07,feiguin07,casula08,hm10a}, this phase  occupies the
center of the trap in a large parameter regime, corresponding in particular 
to the case in which heavy particles are the majority ones. These numerically
exact observations are consistent with the known phase diagrams for 
homogeneous system via a local-density approximation \cite{wang09,orso10,roux11}. 
 
The plan of the paper is the following. First, in Sec.~\ref{sec:model} we discuss typical experimental settings and conditions for two examples:
(i) a heteronuclear Fermi-Fermi mixture such as ${}^{40}$K and ${}^6$Li and (ii) a homonuclear Bose-Bose mixture, where
the mass-imbalance arises due to a spin-dependent optical lattice. This discussion serves to
guide our numerical study.
In Section \ref{sec:fermions}, we provide an extensive analysis of the density profiles of a mass-imbalanced Fermi mixture
and we present a state diagram for the various shell structures. In Sec.~\ref{sec:bosons}, we study both hardcore and softcore
bosons and probe the stability of the dimer and trimer fluid. We conclude each of these section with 
a detailed summary of the  results, while the main aspects    
 are summarized again in Sec.~\ref{sec:summary}. 

\section{Experimental set-ups for mass-imbalanced systems in optical lattices}
\label{sec:model}

The purpose of this section is to describe the experimental set-ups 
to realize mass-imbalanced systems described by the asymmetric Hubbard model. 
In this context, we  discuss heteronuclear mixtures, focussing on the example of the Fermi-Fermi system ${}^{40}$K and ${}^6$Li.
The case of a homonuclear Fermi-Fermi mixture for which the mass-imbalance is induced by spin-dependent  optical lattices  has been described in \cite{liu04}.
Then we turn to homonuclear Bose mixtures in a spin-dependent optical lattice.

\subsection{Model Hamiltonian}
Our numerical analysis will be based on the attractive, asymmetric 1D Hubbard model:
\begin{eqnarray}
H =  & - &  \sum\limits^{L-1}_{l=1, {\sigma}} t_{\sigma}\left(c^\dagger_{l\sigma} c_{l+1\sigma}+h.c.\right)
+ U \sum\limits^L_{l=1} n_{l\uparrow} n_{l\downarrow}\nonumber\\
& + & \sum^{L}_{l=1,\sigma} V_{\sigma}(l-L/2)^2n_l,
\label{one}
\end{eqnarray}
where $c^\dagger_{\ell\sigma}$ creates a fermion with a pseudo-spin $\sigma=\uparrow,\downarrow$ at site $l$, $n_{\ell\sigma}=c^\dagger_{\ell\sigma}c_{\ell \sigma}$, $n_{\ell}=n_{\ell\uparrow}+n_{\ell\downarrow}$ is the local density, $t_{\sigma}$ is the hopping parameter with an explicit dependence on the pseudo-spin index, and $U<0$ is the attractive onsite interaction energy. We set the lattice spacing to unity and impose open boundary conditions. We add a harmonic confining potential parameterized by  constants $V_{\sigma}$. 

Without loss of generality we will consider $t_{\uparrow}>t_{\downarrow} $.
A positive (negative) polarization $P$ corresponds to a majority of light (heavy) particles.
We also introduce the {\it local} magnetization through
\begin{equation}
\langle S_i^z\rangle  = \langle n_{i\uparrow}-n_{i\downarrow}\rangle/2\,.
\end{equation}

We further define the ratio of the trapping potentials as
\begin{equation}
\eta = V_{\uparrow}/V_{\downarrow} \,.
\end{equation} 

\subsection{Heteronuclear Fermi-Fermi mixtures}
\label{sec:fermi_exp}

\begin{figure}
\centering
\includegraphics[width=8cm]{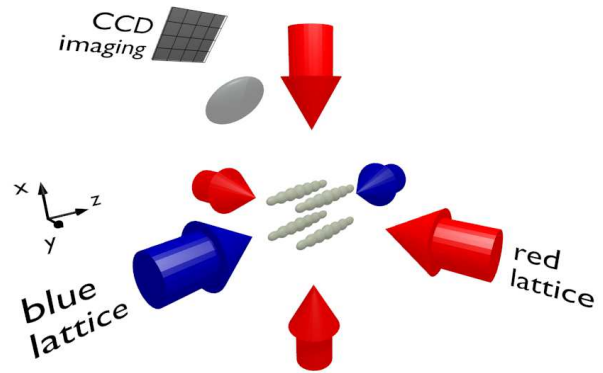}
\caption{\label{fig:setup} An array of independent one-dimensional optical dipole traps is created by intersecting two perpendicular standing waves (red beams). A third orthogonal optical standing wave (blue beams) generates a repulsive lattice potential along the traps. The density profiles of trapped atoms can be observed by an imaging setup from a transverse direction.}
\end{figure}

The proposed experimental setup is schematically shown in Fig.~\ref{fig:setup}. Two pairs of counterpropagating laser beams with orthogonal linear polarization are orthogonally intersected. This creates a 2D array of individual atom traps in the shape of one-dimensional tubes. In this example we choose the wavelength to be $\lambda^{\mathrm{trap}}=1024\,$nm, far red detuned from the atomic transitions. Considering a two-species fermionic mixture of $\Li$ and $\K$, we take into account the transition wavelengths for the D-lines ($671.0\,$nm for lithium and $770.1\,$nm / $767.7\,$nm for potassium). In order to realize one-dimensional optical lattices we superimpose a pair counterpropagating laser beams perpendicular trapping beams and blue detuned with respect to the  wavelengths of the D lines. In the following subsections, we will discuss the resulting trapping potentials $V_{\sigma}$, the hopping parameters $t_{\sigma}$
and the onsite interaction $U$.

\subsubsection{Trapping potentials}

The ratio of the optical trapping potentials $V_{i}^{\mathrm{trap}}$ ($i=$Li,K) is a fixed number that amounts for the given wavelengths to
\begin{equation}
V_{Li}^{\mathrm{trap}}/V_{K}^{\mathrm{trap}}\approx 0.44.
\end{equation}
Note that $V_{\uparrow}=V_{\mathrm{Li}}^{\mathrm{trap}}$ and $V_{\downarrow}=V_{\mathrm{K}}^{\mathrm{trap}}$.
Due to the difference in the optical potentials and the different masses the respective trapping frequencies are generally different:
\begin{equation}
\omega^{\mathrm{trap}}_i = \sqrt{V_{0,i}^{\mathrm{trap}}} \sqrt{\frac{2}{m_i}} \frac{\sqrt{2}}{w_l} \,,
\label{eq:omega_trap}
\end{equation}
where $w_l$ is the waist of the trapping beams and $V_{0,i}^{\mathrm{trap}}$ is the trap depth in the center of the crossed beam setup.

In general the wavelength for the trap can be chosen such that the difference in the optical potentials for the two atomic species  compensates the mass difference such that equal trapping frequencies are obtained~\cite{ospelkaus2006}. This so called \textit{magic} wavelength, $\lambda^{\mathrm{trap}}_{\mathrm{M}}$, occurs for the mixture of $\Li$- $\K$ at $799.9\,$nm. 

\subsubsection{The optical lattice}

\begin{figure}
\centering \includegraphics[width=8.5cm]{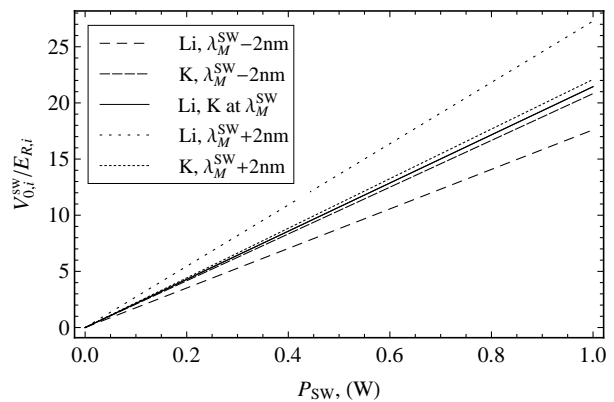}
\caption{\label{fig:Vo} The normalized lattice depth: At the magic wavelength ($\lambda^{\mathrm{sw}}_\mathrm{M}=661.3\,$nm) the lattice depths are equal for $\Li$ and $\K$ (solid line). Two nanometer above (dotted) and below (dashed) the lattice depths differ, while the ratio is inverted at the magic wavelength. Using a beam waist of $200\,\mathrm{\mu m}$  less than one Watt of laser power is sufficient to achieve relevant lattice depths.}
\end{figure}

The one-dimensional optical lattice is realized by superimposing a pair of counterpropagating laser beams perpendicular to the beams that form the trapping array. For a given intensity of the lasers creating the optical lattice, potassium and lithium will experience optical lattice potentials of a different depths as illustrated in Fig.~\ref{fig:Vo}. The experimentally controllable parameters are the amplitudes of the standing wave forming the optical lattice, {\it i.e.}, the lattice depths $V_{0,i}^{\mathrm{sw}}$. These, and the masses determine the respective frequencies:
\begin{equation}
\omega^{sw}_i = \sqrt{V_{0,i}^{\mathrm{sw}}} \sqrt{\frac{2}{m_i}} k 
\label{eq:opt}
\end{equation}

It is useful to introduce the atomic recoil energies as
\begin{equation}
E_{R,i}=\frac{\hbar^2 k^2}{2 m_i}=\frac{\hbar^2\pi^2}{2m_ia^2},
\label{eq:recoil}
\end{equation}
where $k=2\pi/\lambda^{\mathrm{sw}}$ is the wave vector of the standing wave, $\lambda^{\mathrm{sw}}$ is the laser wavelength, and $a=\lambda^{\mathrm{sw}}/2$ is the lattice spacing.
This allows us to express the frequencies as:
\begin{equation}
\omega^{sw}_i = \sqrt{\frac{4E_{R,i}V_{0,i}^{\mathrm{sw}}}{\hbar ^2}}
\end{equation}
For wavelengths of the lattice laser blue detuned  to the atomic transitions there exists a single {\it magic} wavelength, $\lambda^{\mathrm{sw}}_{\text{M}}$, at which the lattice depths $V_{0,i}^{\mathrm{sw}}$ normalized to the respective recoil energies $E_{R,i}$ are equal:
\[
\frac{V_{0,i}^{\mathrm{sw}}}{E_{R,i}} = \tilde{V_0}.
\]
For the $\Li$-$\K$ mixture this occurs at $661.3\,$nm. 
In Table \ref{tab:magicwave} we summarize the magic wavelengths and the ratios of the optical trapping potentials at $\lambda^{\mathrm{trap}}=1024\,$nm for different combinations of alkaline atomic species. As is obvious from the table, in all cases, the trapping {\it potentials} differ.
 
\begin{figure}
\centering\includegraphics[width=8.5cm]{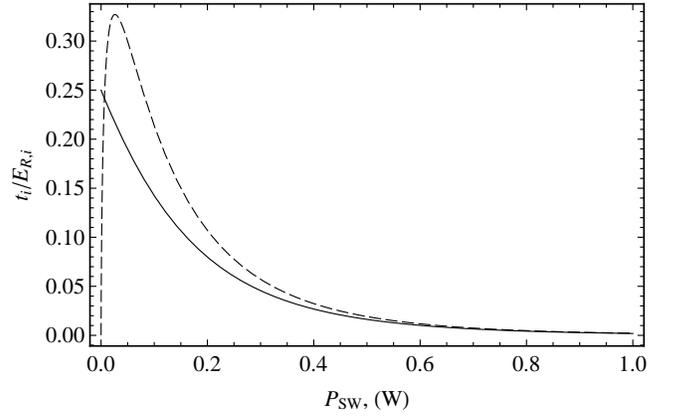}
\caption{\label{fig:tLiK} The tunneling energies at the magic wavelength are the same for $\Li$ and $\K$ (solid line). The asymptotic expression (dashed line) only describes the exact solution of the Mathieu equation at sufficiently high lattice depths (see Eq.~\eqref{eq:t_deep}). The maximum at low lattice depths is not physical.}
\end{figure}

We assume a separable three-dimensional lattice potential of the form
\begin{eqnarray}
V(x,y,z) &=& V_{\parallel}(x)+V_{\perp}(y)+V_{\perp}(z)  \\
& = & V_{0,\parallel} \sin^2(kx) + V_{0,\perp}\left[ \sin^2(ky) + \sin^2(kz) \right]. \nonumber
\end{eqnarray}
This allows us to calculate the hopping matrix elements from the
one-dimensional Mathieu equation. The result is shown in Fig. \ref{fig:tLiK} (solid lines). In the limit of a deep lattice, one can obtain an analytical expression \cite{bloch08}:
\begin{equation}
t_{i,\lambda} =  \frac{4E_{R,i}}{\sqrt{\pi}} \left(\frac{V_{0,i,\lambda}^{sw}}{E_{R,i}}\right)^{3/4}
\mbox{exp}\left(-2\sqrt{\frac{V_{0,i,\lambda}^{sw}}{E_{R,i}}}\right)\quad \lambda=\parallel,\perp,
\label{eq:t_deep}
\end{equation}
with $i=K,Li$ and recoil energies $E_{R,i}$. This result is also included in Fig. \ref{fig:tLiK} (dashed lines), and it agrees with the exact solution for large
$V_0^{\mathrm{sw}}$ or respectively laser power $P_{\mathrm{sw}}$.

\begin{figure}
\centering\includegraphics[width=8.5cm]{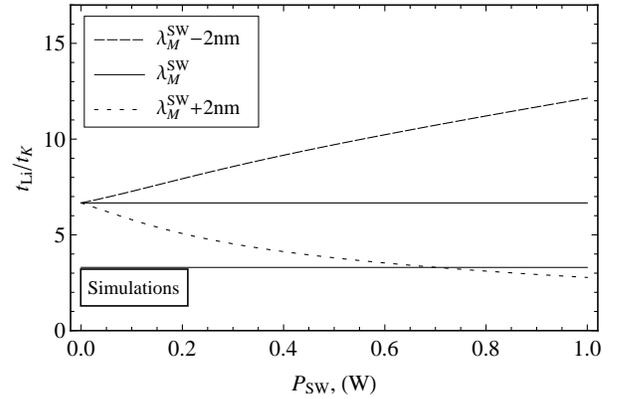}
\caption{\label{fig:tlitk} Ratio of the tunneling energies for $\Li$ and $\K$ for different lattice wavelengths. At the magic wavelength the ratio is constant at the inverse mass ratio (solid line). Two nanometers above and below the magic wave-length the ratio of the tunneling energies differs from the bare mass ratio (dashed lines). By increasing the lattice power the ratio of the tunneling energies can be tuned over a wide range that can be set by the choice of the wavelength. The DMRG and SSES simulations presented in this work are done for a ratio of the tunneling energies  $t_{Li}/t_{K}=3 \frac{1}{3}$.  }
\end{figure}

The ratio of the tunneling energies of $\Li$ and $\K$ is shown in Fig.~\ref{fig:tlitk}.
While the bare mass ratio  is $m_{Li}/m_K\sim 0.15$, we deduce that the ratio of the actual hopping matrix elements
 further depends on the laser power $P_{sw}$  through the lattice depth $V_{0,i}^{\mathrm{sw}}$. This is illustrated in Fig.~\ref{fig:tlitk}  Thus in general, the ratio $t_{\uparrow}/t_{\downarrow}$ can be tuned and is not solely given by the bare mass ratio. In our numerical analysis of a two-component Fermi mixture, we will work at $t_{\downarrow}/t_{\uparrow}=0.3$, corresponding
to $t_{Li}/t_{K}=3$. This requires to detune the wavelength slightly above the magic wavelength $\lambda_{M}^{\mathrm{sw}}$ (compare Fig.~\ref{fig:tlitk}.)

The onsite repulsion $U$ between two atoms of different species is given by \cite{albus03}:

\begin{eqnarray}
U &=& \frac{2\pi \hbar^2 a_{i,j}}{\mu_{i,j}} \int dr^3 |\psi_{i}(x,y,z)|^2 |\psi_{j}(x,y,z)|^2 
\end{eqnarray}
In the expression, $\psi_{\sigma}$ are the Wannier functions , $\mu_{i,j}$ is the reduced mass, and $ a_{i,j}$
is the scattering length that could be different for two atomic species, hyperfine flavors, or Bose-Fermi mixtures.

For the evaluation of $U$ we assume a deep lattice $V_{0,i}^{sw}\gg E_{R,i}$ and that the particles occupy the lowest band. We can then approximate the Wannier functions by the ground state functions of a 1D harmonic oscillator:
\begin{eqnarray}
\psi_i(x,y,z) &= & \psi_{i,\parallel}(x)\psi_{i,\perp}(y)\psi_{i,\perp}(z) \nonumber \\ 
\psi_{i,\parallel}(x)   &=& (\pi\sigma_{i\parallel}^2)^{-1/4} \,\mbox{exp}(-x^2/2\sigma_{i\parallel}^2) \nonumber \\
\psi_{i,\perp}(y)   &=& (\pi\sigma_{i\perp}^2)^{-1/4} \,\mbox{exp}(-y^2/2\sigma_{i\perp}^2)\\
\psi_{i,\perp}(z)   &=& (\pi\sigma_{i\perp}^2)^{-1/4} \,\mbox{exp}(-z^2/2\sigma_{i\perp}^2) \nonumber
\end{eqnarray}
with 
\begin{equation}
\sigma_{i,\lambda} = \sqrt{\frac{\hbar}{m_{i} \omega_{i,\lambda}^{sw}} }\quad \lambda=\perp, \parallel
\end{equation}
Notice that for {\it soft-core bosons}, one needs to consider excited states.

A simple and straightforward Gaussian integration yields:
$$
U_{i,j} = \frac{2\pi \hbar^2 a_{i,j}}{\mu_{i,j}} \frac{1}{\pi^{3/2}} \frac{1}{\sqrt{\sigma_{i,\parallel}^2+\sigma_{j,\parallel}^2}} \frac{1}{\sigma_{i,\perp}^2+\sigma_{j,\perp}^2}.
$$
This expression can be simplified if we work at the magic wavelength for the lattice, since for this case we have:
$$
\sigma_{i,\parallel}=\sigma_{j,\parallel} = \frac{a}{\pi\tilde{V}_0^{1/4}}.
$$
Expressing the energies in units of one of the recoil energies, we obtain
\begin{equation}
\frac{U_{i,j}}{E_{R,i}}=\frac{4a_{i,j}a}{\sqrt{2}\pi^{3/2}}\left(1+\frac{m_i}{m_j}\right)\frac{1}{\sigma_{i,\perp}^2+\sigma_{j,\perp}^2}\tilde{V}_0^{1/4}
\end{equation}

In the extreme 1D limit, this entire expression can be simplified even further, since the Wannier functions become Dirac deltas in the transverse direction:
\begin{equation}
\frac{U_{i,j}^{1D}}{E_{R,i}}=\frac{4a_{i,j}a}{\sqrt{2\pi}}\left(1+\frac{m_i}{m_j}\right)\tilde{V}_0^{1/4}
\end{equation}
These equations indicate the recipe for controlling the interactions, which can be done by either changing the lattice depth with lasers, or by tuning the scattering length using Feshbach resonances. Notice that if the lattice is too deep, multi-band processes can occur, but the validity of the one-band approximation is generally satisfied in practice.

\begin{table}
 \caption{\label{tab:magicwave} Magic wavelengths resulting in identical trapping frequencies or lattice potentials for different combinations of alkaline atoms. The calculation takes both D lines for each species into account.}
 \begin{ruledtabular}
 \begin{tabular}{l|rrr}
 & $V^{\text{trap}}_{0,1}/V^{\text{trap}}_{0,2}$  & $\lambda^{\mathrm{trap}}_{\text{M}}$ & $\lambda^{\mathrm{sw}}_{\text{M}}$\\
 & at $1064\,$nm & (nm) & (nm) \\
 \hline
 $^6$Li$^{40}$K & 0.44 & 799.9 & 661.3\\
 \hline
 $^6$Li$^{87}$Rb & 0.4 & 804.8 & 666.3\\
 \hline
 $^6$Li$^{133}$Cs & 0.23 & 905.1 & 666.7\\
 \hline
 $^{23}$Na$^{40}$K & 0.39 & - & 481.1\\
 \hline
 $^{23}$Na$^{87}$Rb & 0.35 & 945.8 & 551.7\\
  \hline
 $^{40}$K$^{87}$Rb & 0.9 & 807.3 & 655.4\\
  \end{tabular}
 \end{ruledtabular}
\end{table}

\subsection{Homonuclear two-component Bose gases in spin-dependent optical lattices}
In the case of bosons the Hubbard Hamiltonian contains as well intraspecies interactions:
 \begin{eqnarray}\label{H_bos}
H =  & - &  \sum_{i, {\sigma}} \left[ t_{\sigma}\left(b^\dagger_{i\sigma} b_{i+1\sigma}+h.c.\right)
+ U_{\sigma\sigma} n_{i,\sigma} \left(n_{i,\sigma}-1\right) \right] \nonumber\\
& + & U \sum\limits^L_{i=1} n_{i\uparrow} n_{i\downarrow}
+  \sum_{i, \sigma} V_{\sigma}(i-L/2)^2n_i.
\label{eq:one}
\end{eqnarray}
 Here $b^{\dagger}_{i\sigma}$ creates a boson of type $\sigma=\uparrow,\downarrow$ on site $i$.
 In this case, mass imbalance can be realized either in heteronuclear mixtures  (such as $^{87}$Rb/$^{41}$K
\cite{catani2008,catani09})
or in hyperfine mixtures in spin-dependent optical lattices (such as in the case of $^{87}$Rb as realized in a number of recent experiments \cite{mckay2010,gadway2011}.
We will focus the following short discussion on the latter example,
 in which the Hubbard parameters can be tuned by the lattice depth, by
 the optical lattice wavelength, and by an interspecies Feshbach resonance.
 We consider $^{87}$Rb in the hyperfine mixture of $|\uparrow\rangle = |F=1, m_F=1\rangle$
 and $|\downarrow\rangle = |F=1, m_F=0\rangle$ hyperfine states, which are collisionally
 stable and for which several Feshbach resonances have been identified \cite{marte02}.
 State-dependent optical lattices can be easily realized when the
 optical lattice wavelength
 is close to the D1 and D2 lines  -- 795 and 780 nm respectively~\cite{grimm00}. We consider a highly anisotropic
 optical lattice, with wavelength $\lambda_{\perp} = 830$ nm and depth $V_{0,\perp}=40~ E_r$
 for the transverse components defining 1D tubes, and wavelength $\lambda_{||}= 784$ nm
 and variable depth $V_{0,||}$ for the longitudinal component along the tubes.
 Here $E_r = \hbar^2k^2_L/(2m)$ is the recoil energy, where
 $k_L = 2\pi/\lambda_{\perp}$ for the transverse components and $k_L = 2\pi/\lambda_{||}$
 for the longitudinal one. As in the previous section, we use the solution of the Mathieu equation to determine the
 hoppings $t_{\uparrow}$ and $t_{\downarrow}$, and we calculate the intraspecies couplings $U_{\uparrow\uparrow}$ and
 $U_{\downarrow \downarrow}$ in the gaussian
 approximation \cite{bloch08}. 
 We find \emph{e.g} that the hopping ratio $t_{\downarrow}/t_{\uparrow}$ spans the interval $0.3-0.1$ when $V_{0,||}$
 goes from $6~E_r$ to $20~E_r$; in the same parameter range, the ratios $U_{\uparrow\uparrow}/t_{\uparrow}$ and
$U_{\downarrow\downarrow}/t_{\downarrow}$
 are above 10, suppressing double occupancy of the same species: in this case an appropriate description of the system is provided by a model of hardcore bosons, on which we will focus our attention later in the paper.
 The interspecies coupling $U$, which would naturally be very strong
 in this parameter interval ($U/t_{\uparrow} > 10$), can be suppressed by exploiting
 the above cited Feshbach resonances.
 

\section{Mass-imbalanced two-component Fermi gases}
\label{sec:fermions}

\subsection{Overview: population- and mass-imbalanced 1D mixtures}
\label{sec:fermi_over}
In this section we study the density profiles and correlations of a mass- and population imbalanced two-component Fermi
gas with attractive interactions in a harmonic trap.  This extends previous studies of population imbalanced mixtures on the one hand and mass-imbalanced systems
on the other hand. 
We shall briefly review the known results.

The phase diagram of a spin-imbalanced two-component Fermi gas with equal masses and attractive interactions can be obtained 
from the Bethe ansatz, both in the continuum \cite{orso07,hu07} and in lattice models \cite{essler-book}. Only recently,
it was rigorously shown that the partially polarized phase is the one-dimensional analogue of the FFLO state by means
of exact numerical methods \cite{feiguin07,batrouni08,tezuka08,rizzi08,casula08}, confirming predictions from mean-field
theory \cite{buzdin83,machida84} and bosonization \cite{yang01}. In 1D, this means that the pair-pair correlation functions are modulated with 
$Q=k_{F\uparrow} -k_{F\downarrow}$ and  decay as a power law, slower than any competing correlation in the two-particle
channel ($k_{F\sigma}=\pi n_{\sigma}$ is the Fermi momentum). The computation of correlations using the Bethe ansatz is generally very difficult, and therefore, the exponents of 
correlations were first calculated by numerically solving the corresponding the Bethe ansatz equations \cite{luescher08}. Very recently,
first analytical results for the FFLO correlations from Bethe ansatz have been presented in \cite{lee11}. In more general models
that incorporate the coupling of fermions to a molecular channel \cite{fuchs04,tokatly04,recati05}, the partially polarized phase is of the FFLO type on
the BCS side only \cite{hm10,baur10}. The fate of the FFLO state in coupled one-dimensional systems has been recently addressed \cite{parish07,luescher08,zhao08,feiguin09,lutchyn11}. 

For spin-imbalanced systems in one dimension, it is by now well-established that   
there is a two-shell structure in a harmonic trap \cite{orso07,hu07,feiguin07,casula08,hm10a}. 
The central shell is always occupied by the partially polarized phase, while in the wings there is a fully paired phase at low polarizations and a fully polarized phase at large polarizations. Thus, there is a critical polarization $P_c$ at which 
the shell structure changes (exactly at $P_c$, the whole system is in a partially polarized phase). 
This was first predicted by applying the local density approximation to the exactly known phase diagram  \cite{orso07,hu07} and then
confirmed in exact numerical simulations  \cite{feiguin07,casula08,hm10a}. 
The  FFLO-type correlations in the partially polarized phase  are stable against the presence of a harmonic trap \cite{feiguin07}. The effect of temperature 
on the shell structure was studied using both Bethe ansatz methods combined with LDA \cite{kakashvili09} and Quantum Monte Carlo simulations \cite{casula08,wolak10}. In a recent experiment with a 3D array of one-dimensional tubes \cite{liao10}, the theoretical predictions
for the shell structure were quantitatively verified \cite{orso07,hu07,kakashvili09}.
For more details, see the review \cite{feiguin11}.

A mass-imbalanced system of attractively interacting fermions is no longer integrable. The phase diagram for the population
and mass-imbalanced case was therefore obtained by using field theory and DMRG calculations \cite{wang09,burovski09,orso10,orso11,roux11}. In Ref.~\cite{wang09},
the main focus was on the partially polarized phase which shrinks as the mass-imbalance increases due to the instability of a strongly
mass imbalanced systems against a collapse for $P<0$ and phase separation for $P>0$ \cite{batrouni09}. The partially polarized phase is, similar to the case of a spin-imbalanced mixture, 
of the 1D FFLO type \cite{wang09}. A complete phase diagram for the case of 
a majority of heavy particles was presented in Ref.~\cite{orso10}, where in particular, the existence of trimer-fluid phases was demonstrated.
These trimer phases can exist if the densities of the two components obey a certain ratio, namely 
\begin{equation}
2n_{\uparrow}=n_{\downarrow}  \label{eq:trimer_cond}\,.
\end{equation}
This is a {\it necessary} condition only; in Ref.~\cite{orso10} it was shown that the trimer gap goes to zero above a critical
density $n_{\downarrow,c}$. Moreover, in the trimer-fluid phase, the trimer correlations decay algebraically \cite{roux11}, while the 
(s-wave) pairing correlations decay exponentially \cite{orso10}.
Note that there are in fact many more stable multimer bound states at other rational ratios of $n_{\uparrow}$ and $n_{\downarrow}$ \cite{burovski09,orso10,roux11},
which we will not study in this work.

As a consequence of the presence of these trimer phases (and phases of liquids of even larger compounds), the partially polarized phase is separated from the vacuum
by an extended line in the $\mu$-$h$ phase diagram \cite{orso10}, in contrast to the population imbalanced system, where this is a critical point \cite{essler-book,orso07,hm10a}.

Another main difference is the broken particle-hole symmetry 
of the the mass- and spin-imbalanced mixture with respect to the spin-imbalanced case only. This has immediate consequences for the shell structure in a harmonic trap. 
To illustrate this, let us resort to the local density approximation (LDA),  
valid if the trapping potential is sufficiently smooth,
 as it is the case with parabolic potentials whose harmonic oscillator length is
 much larger than the lattice spacing or the interparticle distance. In the case
 of a single species and within LDA, the local properties
 of the trapped system can be quantitatively related to those of a bulk system
 whose chemical potential equals the local chemical potential in the trap, namely
 $\mu_i = \mu - V_{\sigma} (i-L/2)^2$. In the case of a two-component mixture the chemical potential
 is generally species dependent, and one can parametrize the two chemical
 potentials as $\mu_{\uparrow} = \mu + h/2$ and $\mu_{\downarrow} = \mu - h/2$.
 The LDA assigns to local regions of the trap the behavior of a bulk system
 with local average chemical potential $\mu_{\sigma} \to \mu_{\sigma i}$, but with fixed
 chemical potential difference (or magnetic field) $h = \mu_{\uparrow} - \mu_{\downarrow}$.

\begin{figure}[h!]
\centering
\includegraphics[width=9cm]{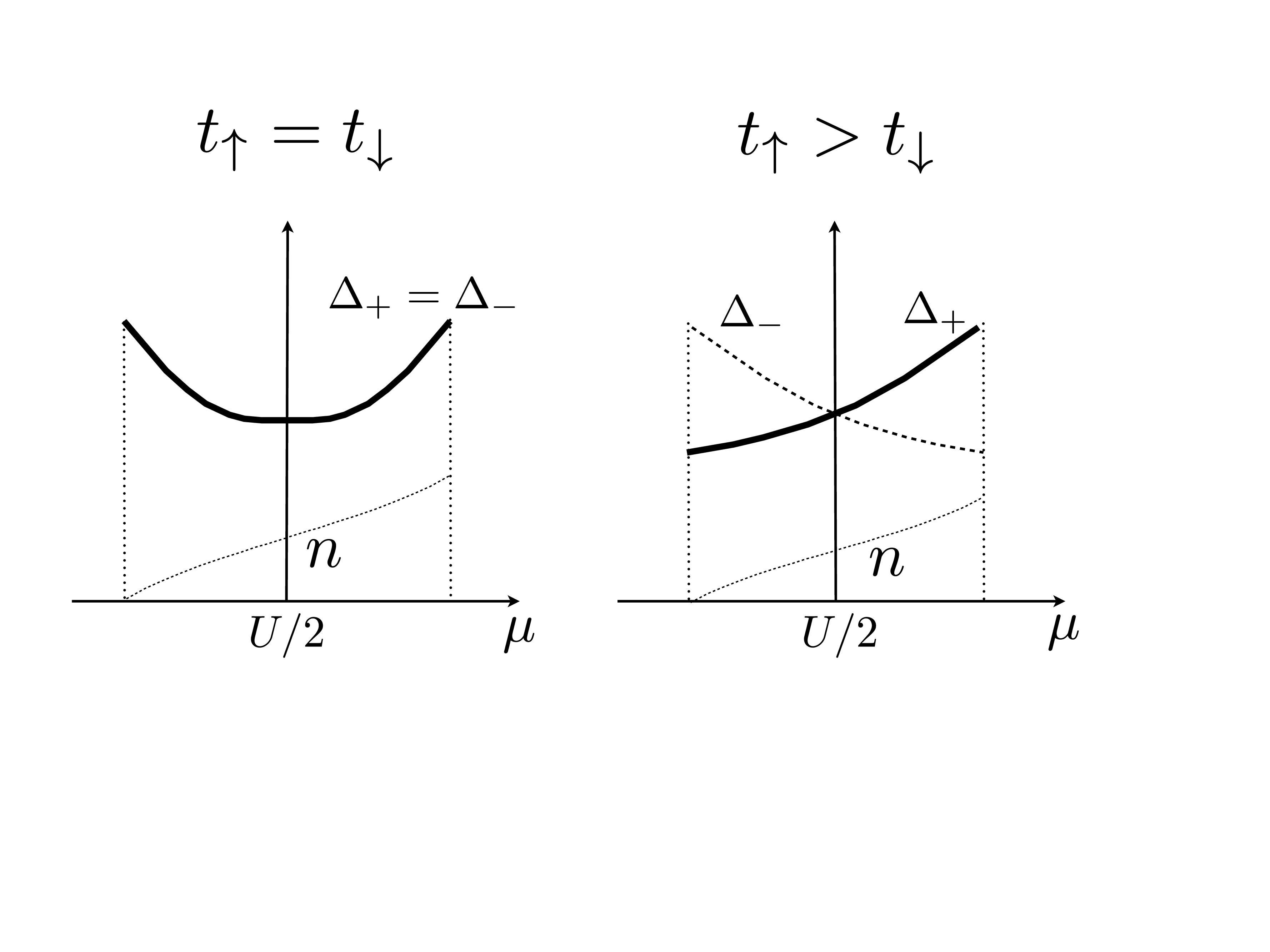}
\caption{(color online) Polarization gaps for the one-dimensional Hubbard model for hardcore
bosons/fermions with equal populations. $n$ is the total particle density. Adapted from Ref.~\cite{wang09}}\label{f.spingap}
\end{figure}

In general, to induce a finite polarization
into the system the magnetic field $h$ has to overcome (in modulus) either one of the following gaps:
\begin{eqnarray}
\Delta_{+} &=& E(N_{\uparrow}+1, N_{\downarrow}-1 ) - E(N_{\uparrow}, N_{\downarrow}) \nonumber \\
 \Delta_{-} &=& E(N_{\uparrow}-1, N_{\downarrow}+1) - E(N_{\uparrow}, N_{\downarrow})~.
\label{eq:spingap}
\end{eqnarray}
where $E(N_{\downarrow}, N_{\uparrow})$ is the ground state energy of a system with a given $N_{\uparrow}$ and $N_{\downarrow}$ in the bulk case. The  gaps $\Delta_{\pm}$ are related to the spin (pairing) gap $\Delta_s$ as
$\Delta_s = \Delta_+ + \Delta_-$.
In other words, to obtain $P\not=0$, one needs either $h > \Delta_{+}$ ($P>0$) or $h < \Delta_-$ ($P<0$).
 In the case of equal masses the polarization gaps
are both equal to $\Delta_s/2$, and hence they are known to be finite through
the Bethe Ansatz solution of the model in the case of attractive interactions $U <0$
\cite{essler-book}. In particular, as depicted in Fig.~\ref{f.spingap}, the Bethe Ansatz
solution predicts that the $\Delta_s$ is a \emph{decreasing} function of $\mu$
below half filling. Therefore, in this situation, applying the LDA to a trapped system
a larger spin gap is found in the trap wings where the density is lower: $h$ being
constant across the trap, this means that a polarization $P$ induced by $h$ will
appear first at the trap center only, moving gradually to the trap wings as $h$
(or $P$) increases.

The situation changes radically in the case of mass imbalance. For a two-component Fermi gas
the gaps $\Delta_{\pm}$ were  calculated via DMRG in Ref.~\cite{wang09} (these results carry over to the case of a two-component gas of hard-core bosons to be discussed later). As could be
anticipated,
the two polarization gaps become unequal, $\Delta_+ \neq \Delta_-$, and, most importantly,
they are no longer particle-hole symmetric functions of the chemical potential
(but the spin gap $\Delta_s$ is). In particular $\Delta_+$ becomes an \emph{increasing}
function of the chemical potential / the density, which is schematically depicted in Fig.~\ref{f.spingap}.
This implies that, in the  mass imbalanced case, an increasing (positive) magnetic field $h$ will first polarize
the trap wings, while the trap center remains unpolarized and exhibits full pairing.

\subsection{Predicted shell structure in a trap} 

Based on the above considerations and the phase diagram of Ref.~\cite{orso10} valid for $t_{\uparrow}/t_{\downarrow}=0.3$, one can expect a variety of shell structures in a harmonic trap using the local density approximation.
For $P<0$, there will be three cases: (i) a partially polarized core plus  fully paired wings for $0<|P|<|P_{c1}|$, (ii)  an extended region with 
only a partially polarized phase at $|P_{c1}|<P<|P_{c2}|$, and (iii) a partially polarized core and fully polarized wings at $|P|>|P_{c2}|$.
 In Ref.~\cite{roux11}, it was demonstrated that 
at $P=-1/3$ there is only a partially polarized phase where the commensurability condition for the existence of trimers 
is fulfilled locally at every point in the trap, confirming the predictions of Ref.~\cite{orso10}.
In the case of $P>0$, we expect only one critical polarization $P_{c3}$, separating a three-shell structure from a two shell structure.
At small $P$, the equal-density phase occupies the core of the system, followed by an intermediate shell that is partially polarized, and
a fully polarized region in the wings. At large $P>P_{c3}$, the core is partially polarized, surrounded by a fully polarized shell.
One goal of our work is to confirm these predictions by using numerically exact DMRG simulations \cite{white92b,schollwoeck05}.

We first compute the density and spin-density profiles at zero temperature using DMRG to obtain
the state diagram of such a system and show that this is consistent with the qualitative 
expectations drawn from the grand-canonical phase diagram \cite{wang09,orso10}. We devote particular attention to
the stability of phases with commensurate densities, i.e., the fully paired phase and the trimer phase against the presence of the trap. 
In addition, we calculate the pairing correlations in the presence of the trap and show that 
they are of the FFLO type in the partially polarized phase. 
Next, we investigate the 
effect of varying $U$ on the critical polarization separating different shell structures at $P>0$. Then, we study the behavior
in {\it spin-}dependent traps, motivated by the discussion from Sec.~\ref{sec:fermi_exp}. Finally, we use quantum Monte Carlo simulations
to analyze the effect of a  finite temperature on the shell structure, focussing on the stability of the fully paired phase at $P>0$.

\subsection{Relation to experimental parameters}

Let us first make the connection between the symbols used here and the experimental set-up more
transparent. We envision a ${}^6$Li-${}^{40}$K mixture in an optical lattice. We associate
\begin{eqnarray}
t_{\uparrow} &=& t_{Li} \\
t_{\downarrow} &=& t_{K}\,.
\end{eqnarray}
We will work with $t_{\downarrow} =0.3 t_{\uparrow}$, which implies to go to a wave-length larger than $\lambda_M$ (compare Fig.~\ref{fig:tlitk}).
We avoid the regime of very strong mass imbalance $t_{\downarrow} \ll t_{\uparrow}$ since in that regime, the system is unstable against
phase separation ($P>0$) or a collapse ($P<0$) \cite{batrouni09}. Since the bare mass ratio is $m_{Li}/m_K \approx 0.05$, it is therefore advantageous to 
tune the system to an intermediate ratio of effective masses.
For most of the simulations we will work with $U=-4t_{\uparrow}$ (implying $U\approx -13.3 t_{\downarrow}$).

Furthermore we will consider the ratio of the trapping frequencies,
parameterized by
$\eta = V_{\uparrow}/V_{\downarrow}$
as a free parameter, and we will first discuss $\eta=1$.
The discussion from Sec.~\ref{sec:fermi_exp} suggests that for the ${}^6$Li-${}^{40}$K mixture, $\eta\approx 0.4$ and we will explore this in
our DMRG analysis as well.

\subsection{Density profiles in a harmonic trap at zero temperature}
\label{sec:density}

\begin{centering}
\begin{figure}
\includegraphics[width=85mm]{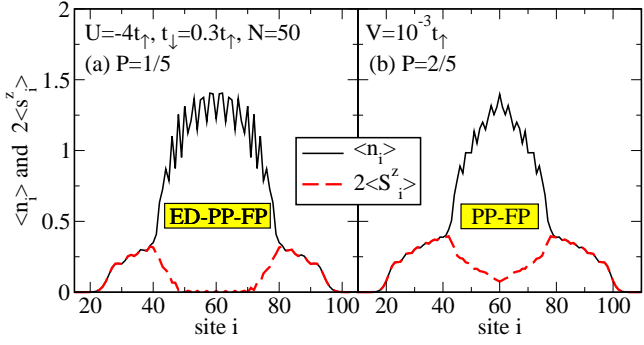}
\caption{
(Color online)
Density profiles for $P>0$: (a) At low $P$,  there are two shells: ED in the core
and PP and FP in the wings. (b) At large $P$, PP sits in the core and FP sits in the wings.
Parameters: $t_{\downarrow}=0.3t_{\uparrow}$, $U=-4t_{\uparrow}$, $N=50$, $V= 10^{-3}t_{\uparrow}$ and (a)  $P=1/5$, (b) $P=2/5$.
(solid lines: $\langle n_i\rangle $, dashed lines: local population difference $-2\langle S_i^z\rangle $).}
\label{fig:p_neg}
\end{figure}
\end{centering}

\begin{centering}
\begin{figure}
\includegraphics[width=85mm]{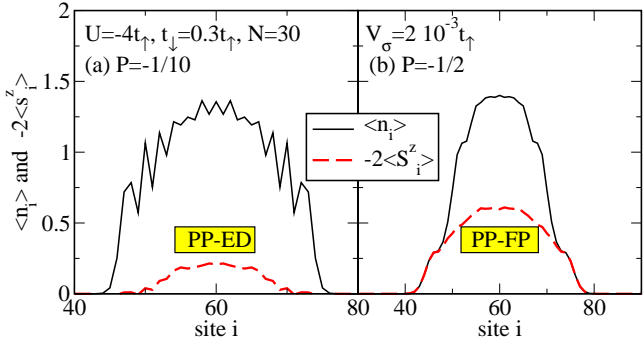}
\caption{
(Color online)
Density profiles for $P<0$: (a) At low $P$,  there are two shells: PP in the core
 and FP in the wings. (b) At large $P$, PP sits in the core and FP sits in the wings.
Parameters: $t_{\downarrow}=0.3t_{\uparrow}$, $U=-4t_{\uparrow}$, $N=50$, $V_{\sigma}=2\,\cdot 10^{-3}t_{\uparrow}$ and (a)  $P=-1/10$, (b) $P=-1/2$.
(solid lines: $\langle n_i\rangle $, dashed lines: local population difference $-2\langle S_i^z\rangle $).}
\label{fig:p_pos}
\end{figure}
\end{centering}

We investigate the density profiles at $T=0$ using DMRG. Our DMRG simulations
are done on a chain with $L=140$ sites, and using
between  $m=400$ and $m=800$ states.
We consider variable trapping potentials, confining the atomic cloud well within the simulation box.
In order to label the emergent shell structures, we introduce a set of
acronyms: PP (partially polarized), ED (equal density), FP (fully polarized), TP (trimer phase).
Then, we use strings of these acronyms that describe the shell structure, going from the core to the wings. As an example, the label PP-ED stands for a system with a PP phase in the core and 
an ED phase in the wings. 

\subsubsection{Equal trapping potentials, $V_{\uparrow}=V_{\downarrow}$, low densities}

In this section, we consider the case of $\eta=1$, i.e., equal trapping potentials for both 
components. Typical density and spin density profiles obtained from DMRG simulations are
shown in Figs.~\ref{fig:p_neg} and \ref{fig:p_pos} for $P>0$ and $P<0$, respectively. 
These results confirm our qualitative 
expectations discussed in Sec.~\ref{sec:fermi_over}, namely that at 
$P>0$, there are three shells for $P<P_{c3}$ (ED-PP-FP) and two at large polarizations $P>P_{c3}$ (PP-FP, compare 
Fig.~\ref{fig:p_neg}).
At $P<0$, there are two critical polarizations separating the different 
shell structures from each other: (i) PP-ED for $P<P_{c1}$,
(ii) PP for $P_{c1}< P< P_{c2} $, and (iii) PP-FP for $P>P_{c2}$ (compare Fig.~\ref{fig:p_pos}).

\begin{figure}[t]
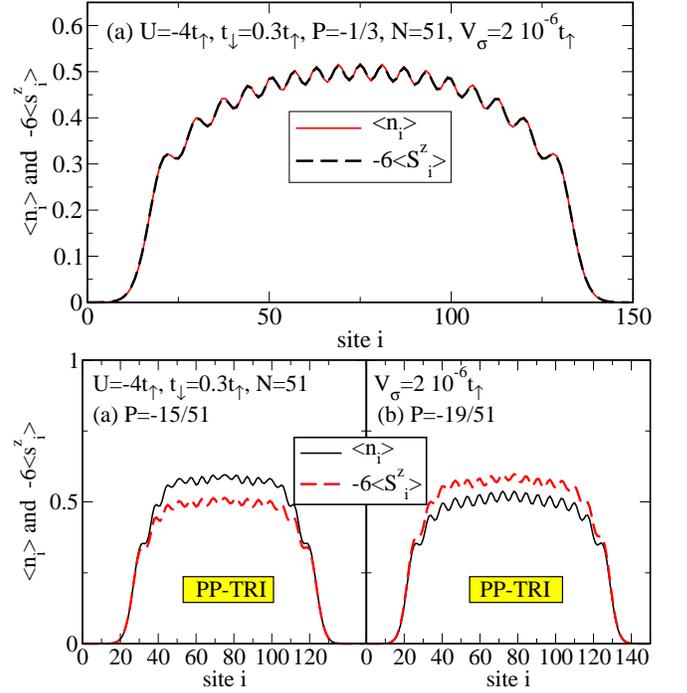

\includegraphics[width=85mm]{Pat_one_third.eps}
\includegraphics[width=85mm]{P_off_one_third.eps}

\caption{
(Color online)
$P<0$, density profiles that match the  trimer condition $n_{\downarrow}=2n_{\uparrow}$ (solid lines: $\langle n_i\rangle $,
dashed lines: local population difference $-6\langle S_i^z\rangle $).
(a) $P=-1/3$: the trimer condition is fulfilled in the entire trap (first shown in Ref.~\cite{roux11}).
(b) $P=-5/21>-1/3 $, (c) $P=-19/51<-1/3$: PP in the core, the trimer condition is fulfilled in a thin outer shell.
Parameters: $t_{\downarrow}=0.3t_{\uparrow}$, $U=-4t_{\uparrow}$, $N=51$, $V_{\sigma}=2\, 10^{-6}t_{\uparrow}$.
}
\label{fig:p_trim}
\end{figure}

The particular case of $ P_{c1} < P < P_{c2} $ is addressed in Fig.~\ref{fig:p_trim}. 
Right at $P=-1/3$ and at sufficiently low density, the trimer condition 
$2n_{\uparrow} =  n_{\downarrow}$ is fulfilled in the entire trapped cloud (see Fig.~\ref{fig:p_trim}(a)
where we reproduce the results from Ref.~\cite{roux11}).
In this parameter regime, the results of Refs.~\cite{orso10} and \cite{roux11} suggest that the ground state
is a trimer fluid. Upon deviating only slightly from $P=-1/3$ (which ensures the trimer condition {\it globally}), 
the trimer condition is no longer fulfilled {\it locally}, i.e. $2\langle n_{i\downarrow}\rangle \not= 
\langle n_{i\uparrow}\rangle$. This demonstrates that in order to stabilize the trimer fluid in a harmonic trap
one needs to fine-tune the global polarization to $P=-1/3$.

\subsubsection{Large density regime}

\begin{figure}[t]
\includegraphics[width=85mm]{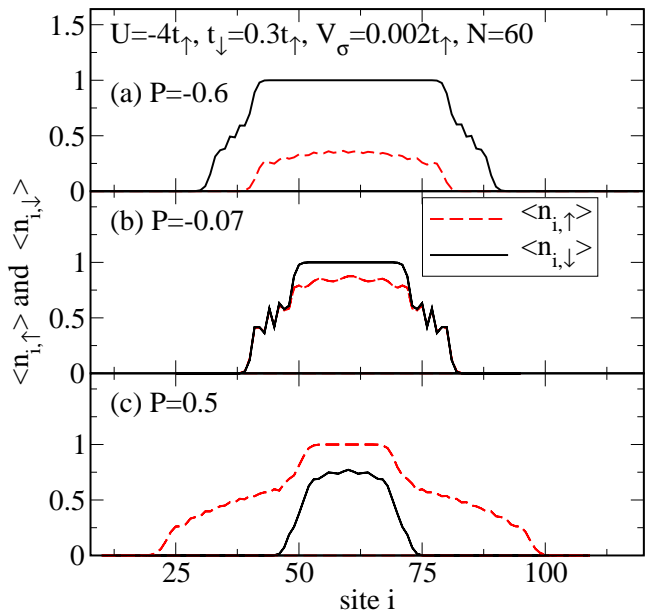}

\caption{
(Color online)
Equal trapping potentials $V_{\uparrow}= V_{\downarrow}$: Emergence of single species band insulators at large densities.
Typical density profiles at (a) $P=-0.6$, (b) $P=-0.07$, (c) $P=0.5$ ($V_{\downarrow}=10^{-3}t_{\uparrow} $, $U=-4t_{\uparrow}$, $t_{\downarrow}=0.3t_{\uparrow}$).}
\label{fig:bis}
\end{figure}

While our main interest is in the low density regime, we here also include examples of typical density
profiles at large densities, i.e., where one of the two components has density $\langle n_{i\sigma}\rangle =1 $.
These are presented in Fig.~\ref{fig:bis}. The first two examples are for $P<0$ and have $\langle n_{i\downarrow}\rangle=1$
in the core of the system, followed by a thin partially polarized shell. At small $P<0$, the outer wings are fully paired [see Fig.~\ref{fig:bis}(a)]
while at large $P<0$, they are partially polarized [see Fig.~\ref{fig:bis}(b)].
In the regime of $P>0$, first the light atoms  form  a band insulator $\langle n_{i\uparrow}\rangle=1$. In the particular example
shown in Fig.~\ref{fig:bis}(c), the core is surrounded by a thin PP phase and a broad fully polarized wing.
Qualitatively, one-species band insulators with $\langle n_{i\sigma}\rangle=1$ form faster at $P<0$ than at $P>0$, i.e., 
by increasing the total particle number 
or by making the trap tighter. Not surprisingly it is energetically favorable to displace the light atoms ($\sigma=\uparrow$) into the outer regions.

\subsubsection{Equal trapping potentials: State diagram}
\label{sec:stdgram}

\begin{figure}
\includegraphics[width=85mm]{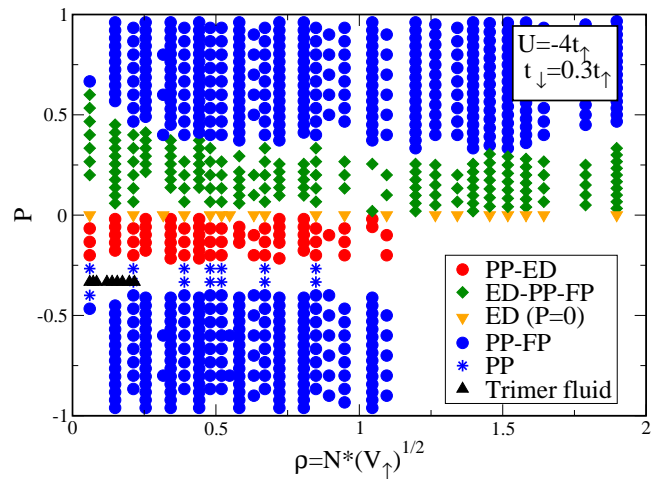}
\caption{
(Color online)
State diagram for $V_{\uparrow}=V_{\downarrow}$ ($U=-4t_{\uparrow}$, $t_{\downarrow}=0.3t_{\uparrow}$).
}
\label{fig:state}
\end{figure}

Our results for the shell structure of a mass- and spin-imbalanced system at $t_{\downarrow} =0.3t_{\uparrow}$ 
and $U=-4t_{\uparrow}$ are summarized in the state diagram Fig.~\ref{fig:state}. We focus on those regions
of polarization $P$ and effective density $\rho=N\sqrt{V}$ in which no component has formed a band insulator yet
(i.e., we restrict ourselves to $\langle n_{i\sigma}\rangle < 1$). Evidently, in most of the state diagram
the partially polarized phase sits in the core with the exception of the region $0<P< P_{c3}$, where the 
equal density phase occupies the central region of the trap. It is interesting to emphasize that this does not
happen for a purely spin-imbalanced system: there, any arbitrarily small polarization pushes the equal density
phase to the outer wings \cite{orso07,hu07,hm10a}. In that sense, the mass-imbalance stabilizes the fully paired/equal density
phase. In particular, if one is interested in this phase, it is therefore not necessary to fine-tune the {\it global} polarization to $P=0$.
This is  a consequence of the broken particle-hole symmetry in a mass-imbalanced system
as discussed in Sec.~\ref{sec:fermi_over}.

The trimer phase, realized at global polarization $P=-1/3$, extends up to $\rho\approx 0.22$ (triangles up in Fig.~\ref{fig:state}). To determine this
point we have followed the density in the center of the trap and compared it to the critical density
of the bulk system beyond which the trimer gap closes \cite{orso10}, which is at $n=0.5$ for $U=-4t_{\uparrow}$. 
In Ref.~\cite{roux11}, it was further shown that at sufficiently low density and at $P=-1/3$, the trimer correlations in the
trap follow a power law. 
One possibility to observe the formation of trimers in an experiment would be to open the
trap and to let the gas expand in 1D in the optical lattice (such an experiment was performed in 2D and 3D optical lattices with
balanced two-component Fermi gases \cite{schneider12}). Due to the formation of heavy objects, i.e., the trimers, that are protected by
the associated excitation gap \cite{orso10}, the expansion should be substantially slower at $P=-1/3$ compared to other 
polarizations. A similar behavior was seen in numerical simulations for the expansion of two-component Fermi gases with a 
high density of doublons \cite{hm09a,langer12}. Our suggestion therefore is to monitor the expansion
velocity as a function of both polarization (at low densities $\rho\lesssim 0.22$) and of effective density at $P=-1/3$. 

Regarding the critical polarizations $P_{ci}$ ($i=1,2,3$), we observe a weak dependence on effective density. Most notably,
going to small $\rho$ further stabilizes the ED-PP-FP  regime. This is consistent with the usual argument
that in one dimension, pairing at equal densities is more robust at low densities \cite{feiguin11}.

It is further instructive to discuss the spatial extent of the atomic cloud, and of the clouds of each individual component.
This follows the analysis of Ref.~\cite{orso07} for the one-dimensional spin-imbalanced Fermi gas, where
it was suggested that the critical polarization can be read off from the polarization dependence of the radii.\
In fact, a similar analysis was then used in the experimental work on density profiles of one-dimensional spin-imbalanced Fermi gases \cite{liao10}.

\begin{figure}
\includegraphics[width=85mm]{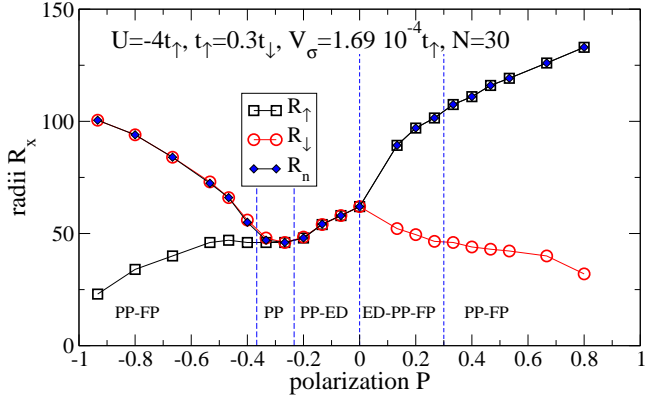}

\caption{
(Color online)
Radius of the particle cloud ($R_n$, triangles), the light ($R_{\uparrow}$, squares), and the heavy fermions ($R_{\downarrow}$, circles)
for $V_{\sigma}=1.69 \, 10^{-4} t_{\uparrow}$ and $N=30$ ($U=-4t_{\uparrow}$, $t_{\downarrow}=0.3t_{\uparrow}$).
The horizontal, dashed lines mark the critical polarizations $P_{c2}$, $P_{c1}$, and $P_{c3}$ (from left to right).}
\label{fig:radii}
\end{figure}

We now  take a cut at a fixed $\rho=N\sqrt{V}$
through the state diagram and we estimate the extent $R_{x}$ ($x=n,\uparrow,\downarrow$) as the region with a finite particle
density, where $R_n$ is the cloud radius and $R_{\sigma}$ are the radii for $\sigma=\uparrow,\downarrow$.
Typical results are shown in Fig.~\ref{fig:radii}. Several aspects deserve being mentioned. First, the 
cloud is the smallest in the region $P_{c2}<P<P_{c1}$, where only a partially polarized phase is present.
Secondly, for $P<P_{c2}$ (where $P_{c2}$ is the critical polarization separating PP from PP-FP), 
one observes $R_{\uparrow} <R_{\downarrow}$, while for $P_{c2}<P<0$, one has $R_{\uparrow} =R_{\downarrow}=R_n$.
At $P_{c1}$ and $P_{c3}$, however, there are no clear features in the dependence of the radii on
polarization (we have to keep in mind, though, that the particle numbers are fairly small).

\begin{figure}
\includegraphics[width=85mm]{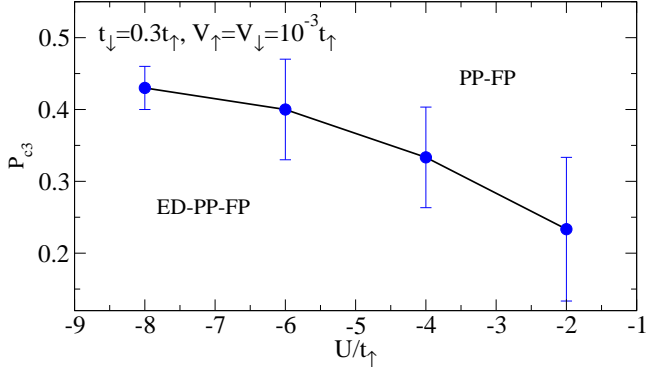}

\caption{
(Color online)
Critical polarization $P_{c3}$ separating ED-FP-PP from PP-FP at $P>0$ as a function of $U$ for  $V_{\uparrow}=V_{\downarrow}$ ($t_{\downarrow}=0.3t_{\uparrow}$, $N=30$, $V_{\sigma}=10^{-3}t_{\uparrow}$).}
\label{fig:pcrit_U}
\end{figure}

While we have mostly focussed on $U=-4t_{\uparrow}$, it is also interesting to study the effect of $U$
on the stability of the ED phase. The dependence of $P_{c3}$ on $U$ is shown in Fig.~\ref{fig:pcrit_U}.
Increasing $|U|$ makes the regime in which the ED phase occupies the core of the system larger, as expected.

\subsubsection{Equal trapping potentials: FFLO correlations in the PP phase}
\label{sec:fflo}
We now show that in the PP phase the FFLO correlations are robust to the presence of the trap. We focus on the region $P<0$, since
at $P>0$, the PP phase sits in the core only for $P>P_{c3}$. In that regime, however, the spin density is typically strongly varying
across the trap, which disfavors clear signatures of the FFLO state. In a harmonic trap, the signatures of the FFLO
state are the cleanest whenever the polarization varies slowly with the local chemical potential \cite{parish07,feiguin07}.

In order to demonstrate the presence of FFLO correlations we compute the Fourier transform of the pair-pair correlations
\begin{equation}
n_k^{\mathrm{pair}} = (1/L)\sum_{lm} \mbox{exp}[ik(l-m)]\, \rho_{lm}\,. \label{nkpair}
\end{equation}
where
$$
\rho_{lm} = \langle c^\dagger_{l\uparrow}c^\dagger_{l\downarrow} c_{m\downarrow}c_{m\uparrow}\rangle \,.
$$
Here, and throughout the paper,  we  adopt the same discretization of the $k$ vectors as  in a homogeneous system,
namely $n_{\sigma}(k=0)$ expresses the number of particles with momentum $k$
in the interval $[-\pi/L,\pi/L)$ for a system of size $L$.
 
\begin{figure}
\includegraphics[width=85mm]{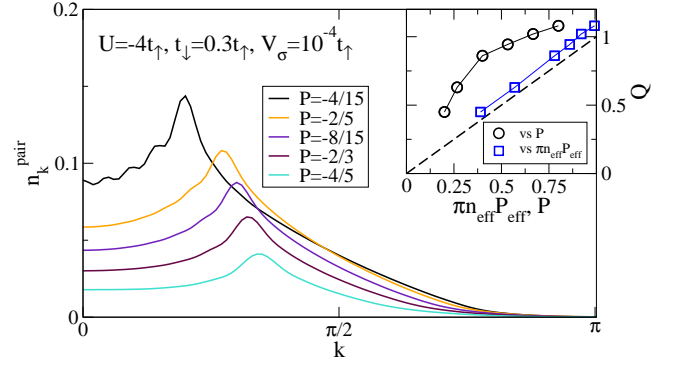}

\caption{
(Color online)
Momentum distribution of pairs in the PP phase at $P<0$ for $V_{\uparrow}=V_{\downarrow}= 10^{-4}t_{\uparrow}$, $U=-4t_{\uparrow}$ ($t_{\downarrow}=0.3t_{\uparrow}$.). We display data for $P=-4/15,-2/5,-8/15,-2/3$, with the peak in $n_k^{\mathrm{pair}}$ shifting from left to right, respectively (we do not
see a clear signature for $0<|P|<1/5$). Inset: Position $Q$ of the maximum in $n_k^{\mathrm{pair}}$ vs $P$ (squares) and vs $\pi n_{\mathrm{eff}} P_{\mathrm{eff}}$ (circles, see text in Sec.~\ref{sec:fflo}). The dashed line is $Q=\pi n_{\mathrm{eff}} P_{\mathrm{eff}}$, i.e., Eq.~\eqref{eq:qeff}.
}
\label{fig:fflo}
\end{figure}

Our DMRG results are displayed in Fig.~\ref{fig:fflo}. Clearly, in the curves with $|P|>0$, we observe
a maximum in $n_k^{\mathrm{pair}}$ at some incommensurate momentum $Q$, indicative of oscillating pair
correlations. In a homogeneous system in one dimension and in the FFLO state, the modulation Q is given by
\begin{equation}
Q = k_{F\uparrow}-k_{F\downarrow} = \pi (N_{\uparrow} -N_{\downarrow})/L\,.
\label{eq:Q}
\end{equation}
i.e., $Q\propto P$.
In a harmonic trap, first of all, not all majority fermions participate in the FFLO state since all particles
in the FP regions have to be excluded. Second, the length entering in Eq.~\eqref{eq:Q} is the one that is actually
occupied by the quasi-condensate. Therefore, $Q$ is {\it not} simply proportional to the global polarization. If one accounts for that, following the procedure described in Ref.~\cite{feiguin07}
then one obtains
\begin{equation}
Q= \pi n_{\mathrm{eff}} P_{\mathrm{eff}} \label{eq:qeff}
\end{equation}
where $n_{\mathrm{eff}}=N_{\mathrm{eff}}/L_{\mathrm{eff}}$ and $P_{\mathrm{eff}}=(N_{\uparrow,{\mathrm{eff}}}-N_{\downarrow,\mathrm{eff}})/N_{\mathrm{eff}}$.
$L_{{\mathrm{eff}}}$ is the region occupied by the quasi-condensate, and all $N_{\sigma,{\mathrm{eff}}}$ are obtained by integrating $\langle n_{i\sigma}\rangle $ over that region.
Equation~\eqref{eq:qeff} should properly describe the scaling of the position $Q$ of the maximum in $n_k^{\mathrm{pair}}$ 
if the spin-density is approximately constant in the PP shell.
We plot $Q$ vs both the total polarization $P$ (circles) and  $\pi n_{\mathrm{eff}} P_{\mathrm{eff}}$ (squares) in the inset of Fig.~\ref{fig:fflo}.
$Q$ deviates from Eq.~\eqref{eq:qeff} by up to 10\%.
This can partially be explained by taking into account the uncertainties of determining $Q$ and $L_{\mathrm{eff}}$ from the finite-size data, and
by finite-size effects.

To complete the discussion, one would need to show that the FFLO correlations decay algebraically along the trap (see the discussion
in Ref.~\cite{feiguin07} for the spin-imbalanced case). For a homogeneous  system with mass- and spin-imbalance, 
the momentum distribution functions were discussed in Refs.~\cite{batrouni09,wang09}, while in Refs.~\cite{orso10,roux11}, it was shown that away from $P=-1/3$,
the (s-wave) paring correlations decay algebraically, modulated with $\cos(Qx)$.

\subsubsection{Unequal trapping potentials, $V_{\uparrow}\not =V_{\downarrow}$}

We now turn to the case of unequal trapping potentials, $\eta\not= 1$.
$\eta>1$ implies that the light atoms are squeezed into the core of the trapped gas whereas for $\eta<1$, they
are being pushed out.

{\it Zero global polarization --} In the case of $\eta\not =1$, a partially polarized phases can be induced if the 
trapping potentials are sufficiently different form each other (results not shown here). However, drastically different trapping potentials are necessary even at small $|U|\sim t_{\uparrow}$ and
the partially polarized phase has a strongly varying spin density. This disfavors this set-up as a way of 
realizing the FFLO state, which leaves clearer fingerprints whenever the spin density varies slowly with the 
chemical potential \cite{parish07}. 

\begin{figure}
\includegraphics[width=85mm]{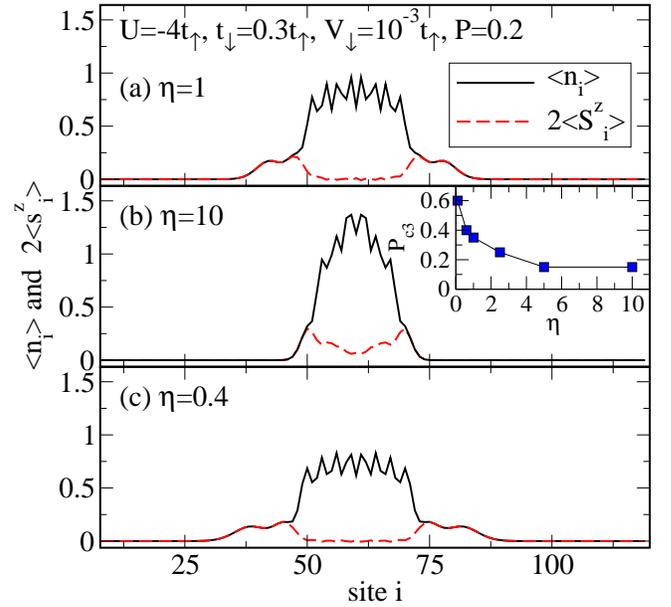}

\caption{
(Color online)
Unequal trapping potentials $V_{\uparrow}\not= V_{\downarrow}$, $P=0.2>0$: 
Typical density profiles at (a) $\eta=1$, (b) $\eta=10$, (c) $\eta=0.4$ ($V_{\downarrow}=10^{-3}t_{\uparrow} $, $U=-4t_{\uparrow}$, $t_{\downarrow}=0.3t_{\uparrow}$).
Inset in (b): critical polarization $P_c$ separating {ED-PP-FP} from {PP-FP} as a function of $\eta$
for $U=-4t_{\uparrow}$, $t_{\downarrow}=0.3t_{\uparrow}$, $N=30$.}
\label{fig:v_unequal}
\end{figure}

{\it $P>0$, stability of the ED phase -- } Next we address the question whether 
unequal trapping potentials drastically alter the state diagram or not.
As an example, we consider $P=0.2$, shown in Figs.~\ref{fig:v_unequal}(a)-(c) for $\eta=1,10$ and $\eta=0.4$, respectively.
At $\eta=1$, the ED phase occupies the core of the system.
Only a very large $\eta$ achieves enough compression of the light particles as to wash out completely the ED phase.
At $\eta = 0.4$ on the other hand, the central ED shell grows bigger 
than for $\eta =1$, as more excess $\uparrow$-particles migrate towards the wings.

To render this observation more quantitative, we plot the critical polarization $P_{c3}$ as a function
of $\eta$ in the inset of Fig.~\ref{fig:v_unequal}(b). $P_{c3}$ monotonously decreases with increasing $\eta$, with a 
very weak dependence on $\eta$ beyond $\eta\approx 5$.  It is important to stress that for the experimental
parameters sketched in Sec.~\ref{sec:fermi_exp}, the equal density shell is actually stabilized since $\eta <1$.

\begin{figure}

\includegraphics[width=85mm]{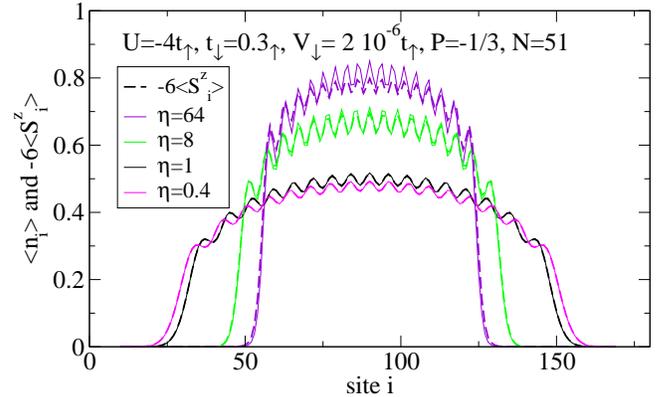}

\caption{
(Color online)
Unequal trapping potentials $V_{\uparrow}\not= V_{\downarrow}$, $P=-1/30$:
Typical density profiles at  $\eta=64,8,1,0.4$ ($V_{\downarrow}=2\cdot 10^{-6}t_{\uparrow} $, $U=-4t_{\uparrow}$, $t_{\downarrow}=0.3t_{\uparrow}$.)
At very large $\eta>1$, the trimer conditions is not fulfilled in the center of the trap, i.e., $\langle n_i \rangle \not= -6\langle S_i^z\rangle$.
}
\label{fig:trimer_eta}
\end{figure}

The trimer phase, realized at global polarization $P=-1/3$ in the entire trap at sufficiently small 
densities, is remarkably stable against varying $\eta$. This is demonstrated in Fig.~\ref{fig:trimer_eta},
where, starting from the parameters of Fig.~\ref{fig:p_trim}(a), we vary $\eta$ from $\eta =64$ to $\eta =0.4$.
By compressing the trap for the light fermions ($\eta>1$) the cloud size shrinks drastically, thus increasing the density in the
center of the trap. Eventually, the trimer condition Eq.~\eqref{eq:trimer_cond} is no longer
fulfilled along the entire trap: the heavy fermions prefer to stay in the wings [see the example of $\eta=64$ in Fig.~\ref{fig:trimer_eta}(a)].
One also drives the core of the system out of the regime in which the trimer gap is finite (compare Refs.~\cite{orso10,roux11}). In the opposite regime,  $\eta<1$, the cloud expands, yet down to $\eta \sim 0.1$, we still observe a perfect match
of $\langle n_i\rangle =-6 \langle S^z_i\rangle $, equivalent to $2 \langle n_{i\uparrow}\rangle = \langle n_{i\downarrow}\rangle $.

\begin{figure}
\centering
\includegraphics[width=9cm]{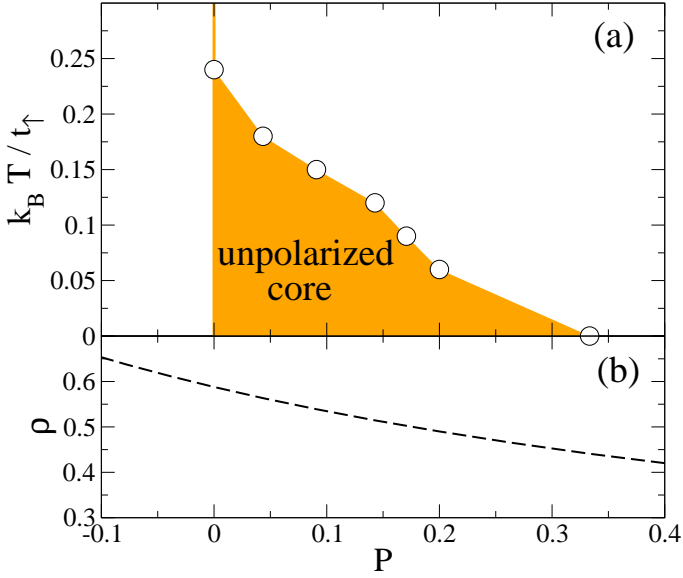}
\caption{\label{fig:finite-T}(color online) (a) Finite-temperature state diagram of a mass imbalanced mixture with 
$t_{\downarrow} = 0.3 t_{\uparrow}$, $U = -4t_{\uparrow}$, and $V_{\uparrow}=V_{\downarrow} = 1.5 \cdot 10^{-4} t_{\uparrow}$. The polarization scan is obtained
by fixing $N_{\uparrow} = 24$ and varying $N_{\downarrow}$. (b) Trajectory in the $P-\rho$ plane (compare Fig.~\ref{fig:state}) described with the
above parameters. }
\end{figure}

\subsection{Density profiles at finite temperatures}

As seen in the previous sections, the formation of an ED dimer liquid in the trap center is very robust to the presence of a finite, positive polarization $P$
for a large window of characteristic densities $\rho$. 
Here we probe the robustness of this phenomenon to the further effect of finite temperatures, by making use 
of quantum Monte Carlo (QMC) simulations based on the Stochastic Series Expansion (SSE) algorithm 
\cite{syljuasen02}. In our present study we use a canonical formulation based on double directed-loop updates \cite{roscilde08}. 
Fig.~\ref{fig:finite-T} refers to a path through the state diagram of Fig.~\ref{fig:state}. This corresponds to an the  experimentally meaningful situation, in which
$N_{\uparrow}=24$ is kept fixed, while $N_{\downarrow}$ is varied, scanning the $P$ axis at variable $\rho$. The actual trajectory in the $P-\rho$ plane is indicated in the figure. Here 
$t_{\downarrow} = 0.3 t_{\uparrow}$ and $U = -4 t_{\uparrow}$ as in the previous sections, 
while $V_{\uparrow} = V_{\downarrow} = 1.5\cdot 10^{-4} t_{\uparrow}$.  As a criterion for dimer liquid formation in the trap center we require that the local magnetization vanishes over the 10 central sites. This criterion is met at $T\approx 0$ (actually $T = 4\cdot 10^{-3} t_{\uparrow}$, ensuring the elimination of thermal effects) over a sizable polarization range $0 \leq P \lesssim 1/3$, 
as already seen in Fig.~\ref{fig:state}. As $T$ is increased above zero, the polarization range featuring an unpolarized dimer liquid (i.e., the ED phase) in the trap core shrinks, but it remains  
sizable up to temperatures $T \approx 0.2 ~t_{\uparrow}/k_B$, proving the robustness of ED pairing to realistic conditions. Indeed, even though pairing correlations in one-dimensional systems become short-ranged as soon as $T$ becomes finite, the finite gaps $\Delta_{\pm}$ 
that were defined in Eq.~\eqref{eq:spingap}
 prevent the thermally excited majority ($\uparrow$) particles from flowing from the trap wings into the center. 
The parameter region featuring an unpolarized core in Fig.~\ref{fig:finite-T} shrinks asymmetrically from the side of positive 
polarizations: indeed at higher $P$ the excess $\uparrow$ particles sitting in the trap wings have a higher potential energy, and therefore they need a smaller thermal energy to overcome the gaps $\Delta_{\pm}$ and to flow into the trap center.

\subsection{Heteronuclear, mass-imbalanced Fermi gases: Summary}

So far we have studied the properties of three pairing states of a spin- and
population imbalanced Fermi gas in a harmonic trap, namely the equal density 
(or dimer fluid) phase, the trimer fluid phase and the FFLO state.

Our main result for the ED phase is that it occupies the core of a trapped
system over a wide range of positive polarizations.
Unequal trapping potentials which are typical for a heteronuclear system destabilize
the ED phase if the confinement for the light particles is much stronger than 
for the heavy ones, whereas in the opposite regime, the ED phase is favored.
This phase is stable against thermal fluctuations as long as they do not overcome
 the polarization gaps $\Delta_{\pm}$.

\section{Mass-imbalanced two-component Bose gases}
\label{sec:bosons}

While fermionic mixtures have been shown to support three different types
of superfluid phases, FFLO, ED and trimer phase, which are stable even in the presence of a
trapping potential, the question naturally arises whether such pairing instabilities
are indeed possible when considering bosonic binary mixtures.
The main difference between fermionic and bosonic statistics resides in the 
fact that bosons can (quasi-)condense, resulting in a sharp peak at wavevector
$k=0$ in the momentum distribution
\begin{equation}
n_{\sigma}(k) = \frac{1}{L} \sum_{ij} e^{i k (r_i-r_j)} B_\sigma(i;j)
\end{equation}
where
\begin{equation}
B_\sigma(i;j)=\langle b^{\dagger}_{\sigma,i}b_{\sigma,j}\rangle
\label{e.Bij}
\end{equation}
is the one-body density matrix (OBDM), and $b_{i\sigma}, b^{\dagger}_{i\sigma}$ are bosonic
operators. 
Strictly speaking, the distinctive feature of phases dominated by one-body coherence
is the power-law decay of the OBDM, Eq.~\eqref{e.Bij} - 
which may or may not result in a divergent $n(k=0)$ peak depending on the 
exponent of the power law. Yet in a trapped system such a power-law behavior
is hard to extract, due to the inhomogeneity imposed by the trap; and, 
more importantly, from an experimental point of view the 
most accessible observable is the momentum distribution. Hence in the following
we will concentrate on the $n_{\sigma}(k=0)$ (or condensate) peak, and in particular on its 
relative changes in height and width as one-body coherence is suppressed
or enhanced. 

The model Hamiltonian in Eq.\ref{H_bos} has been investigated in a series of
regimes. If the densities, the masses and the interaction parameters are the same for the
two species, a paired
phase appears above a critical value of the interaction ratio $|U|/U_{\sigma\sigma}$
\cite{rizzi08b,mathey2007,hu2009},
that is, one needs sufficiently strong intraspecies repulsion in order to open 
a spin gap which suppresses one-body coherence. The resulting paired phase has
an exponentially decaying OBDM, and a power-law decaying pair-correlation function, 
\begin{equation}
P(i;j) = \langle b^{\dagger}_{i{\uparrow}} b^{\dagger}_{i{\downarrow}}  b_{j{\downarrow}} b_{j{\uparrow}}\rangle~.
\end{equation} 
A number of numerical
and analytical calculations support such a picture, which fully recovers the fermionic
case in the Tonks-Girardeau (or hardcore) limit $U_{\sigma\sigma}\rightarrow\infty$. 
In the presence of a finite mass imbalance, the ED paired phase may evolve into a 
charge-density-wave phase and then into a crystalline phase for specific filling fractions,
as explicitly shown in the hardcore case in Ref.~\cite{roscilde2012}. Finally, 
a collapsed phase appears in the strongly attractive regime $|U|/U_{\sigma\sigma}\gtrsim1$,
since the intraspecies repulsion is not sufficiently strong to prevent very large single-site
occupancies.

 On the other hand, in the absence of a trap an arbitrary imbalance
 in the populations will lead to a revival of one-body coherence, which becomes
 algebraically decaying (the same property applies to fermions). In the case of bosons, 
 finite-momentum pairing is not observed in the presence of population imbalance, so that 
 population imbalanced bosonic mixtures with attractive interactions do not feature 
 qualitatively different signatures with respect to mixtures of decoupled species. 
 This issue is quite relevant experimentally, given that having perfectly 
 balanced populations is essentially impossible: even in the case of homonuclear
 mixtures, fluctuations of order $\sqrt{N_{\sigma}}$ are generally expected
 in the population of both species. 
 
 \subsection{Pairing in trapped hardcore bosons with mass and population imbalance} 
 
 Is then bosonic pairing impossible to observe experimentally in one-dimensional
 gases? Luckily the answer is ``no", if one adds the two further ingredients which
 are the central topic of this paper, namely mass imbalance \emph{and} a trapping potential. 
 In the hard-core case, one can extend the LDA argument presented in Sec.~\ref{sec:fermi_over}
 and conclude that, in presence of mass imbalance, a dimer fluid is stable in the center
 of the trap up to a finite polarization, depending on $t_{\downarrow}/t_{\uparrow}$ and $U$.

\begin{figure}
\centering
\includegraphics[width=7cm]{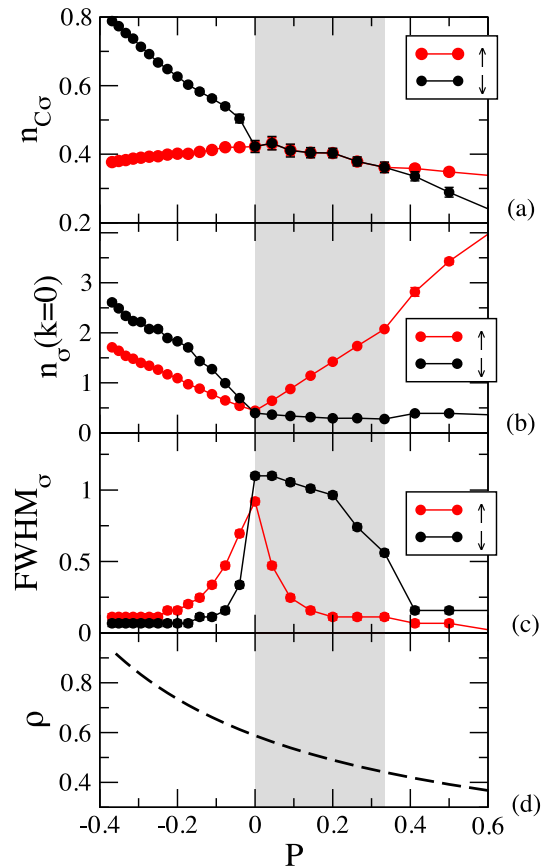}
\caption{\label{fig:T0Nup24}(color online) Polarization dependence of core density and momentum distribution in a mixture with $N_{\uparrow}=24$ and varying $N_{\downarrow}$. The system size $L=140$ enters into the definition of $n_{\sigma}(k=0)$. Other parameters as in Fig.~\ref{fig:finite-T}. (a) Core density; (b) height of the condensate peak; (c) full width at half maximum (FWHM) of the condensate peak (in units of the inverse lattice spacing); (d) trajectory in the $P-\rho$ plane.}
\end{figure}

This picture is fully confirmed by a direct calculation of
trapped hardcore bosons, shown in Fig.~\ref{fig:T0Nup24}. There we simulate
an experimentally meaningful situation of a trapped hardcore boson mixture,  
in which $N_{\uparrow}=24$ is kept fixed, 
while $N_{\downarrow}$ is varied, thereby varying the polarization $P$ continuously. 
We monitor the evolution of the core density $n_{C\sigma}$ (averaged over the 10 core sites)
revealing that a finite window of \emph{positive} polarizations exists for which the 
core remains unpolarized, so that $n_{C\uparrow} = n_{C\downarrow}$. In this 
situation ED pairing is robust at the trap center, and this has strong signatures in the 
\emph{global} coherence properties of the
cloud, captured by the momentum distribution. Indeed in the polarization window in which the core densities are equal, 
the peak height $n_{\downarrow}(k=0)$ is strongly suppressed, and it hardly changes as $P$ decreases, even
though $N_{\downarrow}$ is increasing. The peak height $n_{\uparrow}(k=0)$ shows
two kinks at the two boundaries of the polarization window in question. In particular
the point $P=0$ marks a sharp kink-like minimum associated with full pairing of 
all particles across the trap. Similar features are revealed in the full width at half
maximum (FWHM) of the $n_{\sigma}(k)$ distributions: the polarization window 
with equal core densities shows a strong enhancement of the FWHM for
the $\downarrow$-particles, and its boundaries are marked by two kinks in the FWHM
for the $\uparrow$-particles. Hence bosonic pairing at the trap center is 
robust to the presence of a finite positive polarization, and it bears strong
signatures at the level of the momentum distribution. 

 A detailed calculation of the polarization gap $\Delta_+$ is still missing for 
 the softcore boson case. Nonetheless in the following we will show that
 for realistic parameters a softcore boson mixture in a trap exhibits a similar behavior
 as well. 
 
 \begin{figure}
\centering
\includegraphics[width=7cm]{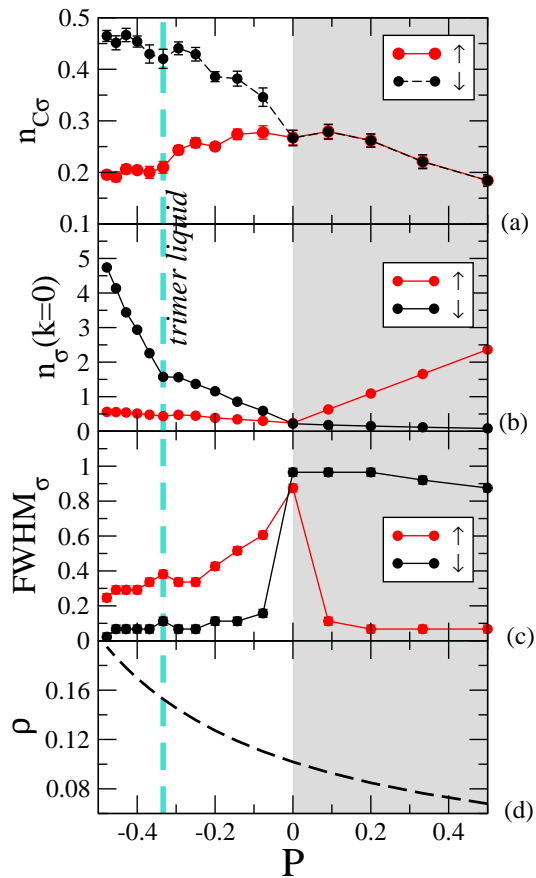}
\caption{\label{fig:trimerformation-nk0}(color online) Evolution of core density and momentum distribution in a mixture with $N_{\uparrow}=12$ and varying $N_{\downarrow}$; here $V_{\uparrow}=V_{\downarrow}=1.8\,\cdot  {10^{-5}} t_{\uparrow}$, and the other parameters are as in Fig.~\ref{fig:T0Nup24}. (a) Core density; (b) height of the condensate peak; (c) full width at half maximum (FWHM) of the condensate peak; (d) trajectory in the $P-\rho$ plane.}
\end{figure}

 \subsection{Trimer formation in hardcore bosonic mixtures: momentum distribution signatures}
 \label{s:trimer_onebcoherence}
 
 In this and all of the following sections, we will focus on the case of further population imbalance, and
 in particular, the case $P < 0$, in which, as discussed earlier, ED pairing is always absent in the trap center. 
In this regime trimer formation can instead appear when $P=-1/3$. 
Such a case of strongly imbalanced population has not been extensively studied
yet; in particular, the emergence of trimer formation has up to now been discussed only in 
systems of fermionic atoms or dipolar molecules \cite{petrov2007,burovski09,orso10,
dalmonte2011}. The key point
here is to qualitatively and quantitatively understand the competition between
possible trimer instabilities and the emergence of either phase separation or
single species superfluidity as a function of the hopping imbalance, which is
expected to play a prominent role as in the fermionic case.

We begin our discussion with the case of hardcore bosons, moving then to the case of softcore bosons in the next sections. 
Ref.~\onlinecite{orso10} has shown that, in the grand-canonical ensemble, a mixture of hardcore bosons or fermions with mass imbalance $t_\downarrow/t_\uparrow = 0.3$ shows a direct transition from a trimer liquid phase with fixed polarization $P=-1/3$ to a vacuum phase as the chemical potential is lowered. Hence, within the LDA approximation, a trapped mixture cannot realize a trimer liquid phase in the trap center unless its polarization is exactly at $P=-1/3$: indeed, if the polarization differs from $-1/3$, the extra particles of either species cannot be accommodated in the trap wings, given that a trimer liquid in the trap center can only be flanked by an empty region.

This condition seems to put a very serious limitation to the possibility of observing a trimer liquid phase in a realistic trapped system, given that, as mentioned above, a fine tuning in the population of the two species is experimentally very hard. Nonetheless, strong signatures of the formation of a trimer liquid phase at $P=-1/3$ can be seen in a broader range of polarizations, and specifically in the momentum distribution. Fig.~\ref{fig:trimerformation-nk0} shows a polarization scan at very low temperature in a mixture containing $N_{\uparrow} = 12$ particles and a variable number $N_{\downarrow}$ of $\downarrow$-particles with $t_\downarrow = 0.3 t_\uparrow$. We chose a very weak trapping potential ($V_\uparrow = V_\downarrow = 1.8*10^{-5} t_{\downarrow}$), leading to low densities in the system, for which trimer formation is a most robust phenomenon \cite{orso10}. We observe that, as expected from LDA, the polarization $-1/3$ is achieved in the trap center ($n_{C\downarrow} = 2 n_{C\uparrow}$) only when the global polarization $P$ is exactly at the same value. When the trimer condition on the polarization is satisfied, a trimer liquid appears to form, as shown by an anomaly in the one-body coherence properties, namely a weak suppression in the height of the condensate peak $n_{\sigma}(k=0)$ as well as a slight enhancement of its width. But the most notable feature is that the formation of trimers at polarization $P=-1/3$ influences the whole evolution of the one-body coherence properties for nearby values of $P$. Indeed a sharp kink appears in the height of the condensate peak $n_{\downarrow}(k=0)$ at $P=-1/3$, 
marking a net change of slope in the dependence of this quantity on $P$. In fact a similar kink is observed for $P=0$, at which the ED paired phase spreads throughout the trap. That kink is associated with the fact that $\downarrow$-particles added to the system to give $P<0$ will not form dimers with $\uparrow$-particles, giving rise to a strong enhancement of $\downarrow$-particle coherence. On the other hand, the further enhancement of coherence for $P < -1/3$ shows that for $-1/3 < P < 0$ the $\downarrow$-particles added to the system have the tendency to form trimer bound states with the $\uparrow$-particles,      
and it is only when the trimer formation has been saturated at $P=-1/3$ that the  one-body coherence grows faster with increasing $N_{\downarrow}$, whence the kink at $P=-1/3$.
 In principle the extra $\downarrow$-particles added to give $P<-1/3$ might form quartets (3 $\downarrow$'s and 1 $\uparrow$), which would lead to an another  kink in the condensate peak with P becomes lower than $-1/2$ (saturation of quartets), and the same for  bound states formed from even more particles. On the other hand the binding energy of these complexes, if finite, will be increasingly low as the size of the complex increases, so that the kinks in the $n_{\downarrow}(k=0)$ curve will be increasingly weak.

\subsection{Trimer binding energy for bosonic mixtures}

In the present section we focus on softcore bosons, and we address the experimentally relevant case of $^{87}$Rb mixtures
confined in species-dependent optical lattices. As discussed in Sec.~\ref{sec:model}, this case is
an archetypical case aimed at underpinning the interplay between bosonic soft-coreness and trimer
pairing. Such settings provide intermediate hopping asymmetry and sufficiently
large intraspecies repulsion, as required to avoid phase separated regimes for
large values of $U/t_\uparrow$. We will start our discussion by considering
the effect of a finite intraspecies interaction in the low-density limit, identifying 
the more favorable parameter setup to observe trimer physics in a finite,
inhomogeneous system, which will then be investigated by means of DMRG
and QMC simulations. 

Understanding the three-body pairing mechanism in the low-density limit 
sheds light onto the more complicated many-body picture: in fact, 
a finite trimer binding energy with magnitude comparable to the hopping rates $t_{\sigma}$
is a good starting point to observe the TP phase at the many-body level, as has
been noticed in the fermionic case \cite{orso10}. At the few-body level,
the main difference between the fermionic case and the bosonic one stems from 
the finite intraspecies repulsion, which, in a $|U|\gg t_{\sigma}$ perturbation picture,
allows for additional exchange processes in the bosonic case. At a qualitative
level, these terms soften the repulsion between a bound $\uparrow-\downarrow$ pair 
and a 
$\downarrow$-particle, thus potentially increasing the trimer binding energy by a factor
of order $\simeq t_\downarrow^2/(U_{\downarrow\downarrow}-U)$.

\begin{figure}[t]
\includegraphics[width=65mm]{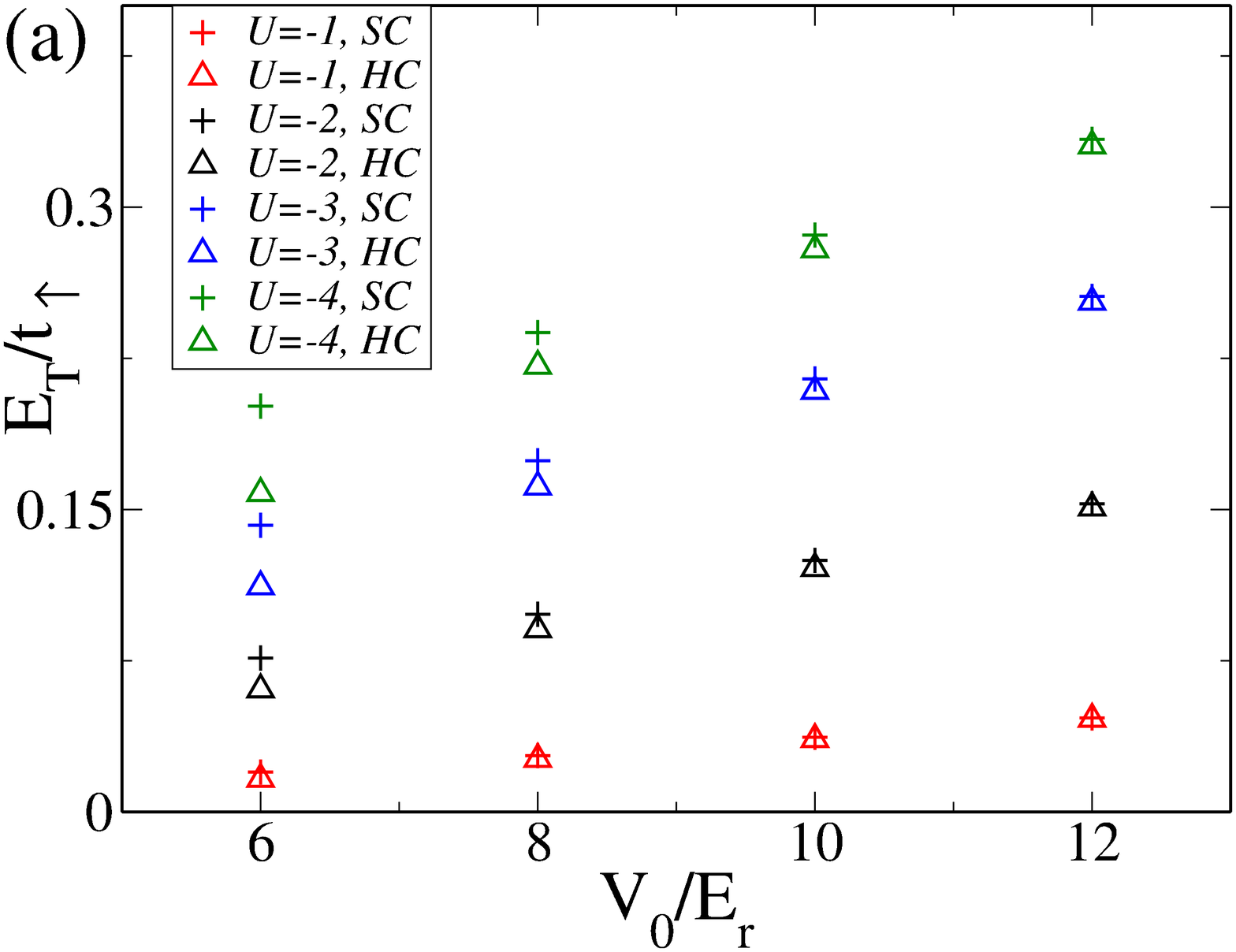}
\includegraphics[width=65mm]{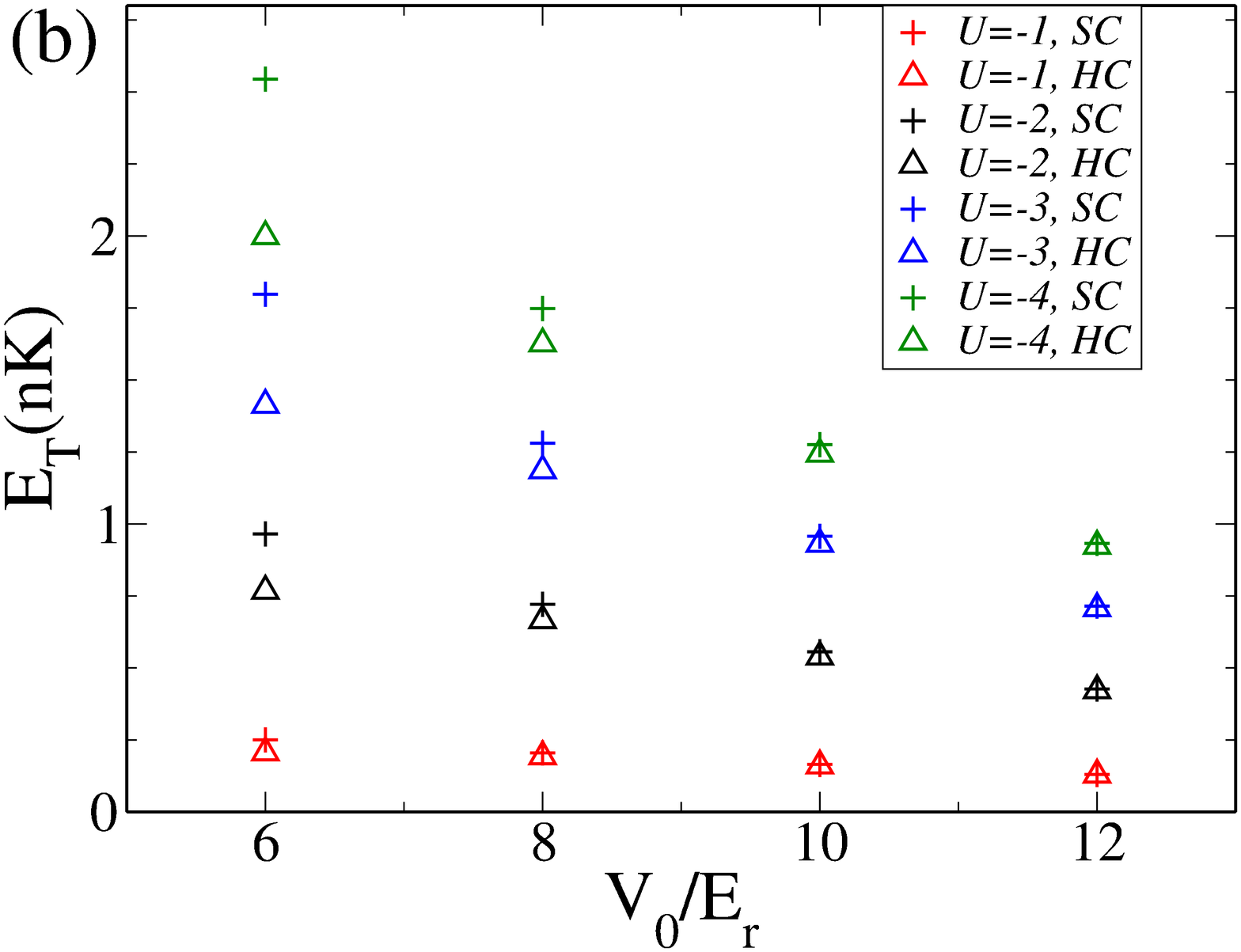}
\caption{Trimer binding energy for $^{87}$Rb mixtures as a function of the lattice depth $V_0$ (in 
units of the recoil energy $E_r$. Note that here, the recoil energy is identical for the two components as it only depends on the bare mass). Panel (a): $E_T$ in units of $t_\uparrow$. In this case, deeper lattices lead to larger mass/ratio, and such a larger $E_T$. Panel (b): $E_T$ in nK units as considered in Sec. \ref{sec:model}.
Deeper lattices strongly reduce the temperature scale, as evident from the sharp decrease of the binding energy in such scale. For all interaction strengths considered, the soft-core case has typical energy scales always larger than the hard-core one.}
\label{f:Et}
\end{figure}

In order to get quantitative information, we evaluate the trimer binding
energy via DMRG simulations, by considering the quantity
\begin{equation}\label{eq:E_T}
E_T(L)=E_L(1,1)+E_L(0,1)-2E_L(1,2)
\end{equation}
where $E_L(N_{\uparrow},N_{\downarrow})$ is the ground state energy
of a system of length $L$ and population $N_\uparrow,N_\downarrow$ respectively.
The trimer binding energy, which corresponds to the trimer gap in the zero-density limit
\cite{orso10}, is then defined as $E_T=\lim_{L\rightarrow\infty}E_T(L)$,
and can be extrapolated using a fourth-order algebraic fit of simulation data 
with $L=20,40,60,80,100$. Typical results for the parameter range $V_0/E_r\in[6,12]$
are shown in Fig.~\ref{f:Et} (upper panel), where, as a comparison mark, we also present
results for the hard-core (HC, $U_{\sigma\sigma}=\infty$) case. The first notable
feature is that the trimer binding energy of soft-core (SC) bosons is always larger
than the HC one, thus confirming the raw qualitative picture described above;
the difference between the two notably shrinks at larger lattice depths, where the 
intraspecies repulsion becomes larger ($U_{\downarrow\downarrow}/t_{\uparrow}\simeq56.9$
for $V_0=12$). Moreover, deeper lattices induce larger mass imbalance, 
which by itself is reflected in a larger binding energy with respect to the hopping
rate $t_{\uparrow}$.

In order to establish  the optimal experimental setting to observe
trimer physics, one has to take into consideration that a finite temperature $T$
in ultracold gases may indeed prevent any relevant observation of paired phases
with associated binding energy smaller that $k_BT$. In the lower panel of Fig.~\ref{f:Et}, we show 
the binding energy in nK units in the same parameter regime as in the upper panel:
in fact, large values of $V_0$ significantly decrease the absolute value of $E_T$,
making the effect of thermal fluctuations more and more relevant.

Combining the aforementioned arguments, we conclude that the best setting
where trimer physics may indeed be observable is determined in our case by
the condition $V_0/E_r \approx 6$; moreover, since in this case $U_{\downarrow\downarrow}
\simeq U_{\uparrow\uparrow}\simeq 10t_{\uparrow}$, we will choose in the following
$U= -3 t_\uparrow$, as larger interspecies attraction may indeed lead to a collapse in an 
inhomogeneous setup.

\subsection{Density profiles and correlation functions in a harmonic trap at 
zero temperature}

In the following, we employ DMRG calculations in order to elucidate the presence
and stability of the trimer phase in trapped systems, keeping up to
400 states per block and applying up to 10 sweeps, with a truncation error
in the final DMRG step usually smaller than $10^{-7}$. We consider
equal trapping potentials, $V_{\sigma}=V$, starting with a commensurate
polarization $P=-1/3$ and then study the effects of incommensurate
polarizations. 

\subsubsection{Equal trapping potentials, commensurate polarization}
\begin{figure}[b]
\includegraphics[width=65mm]{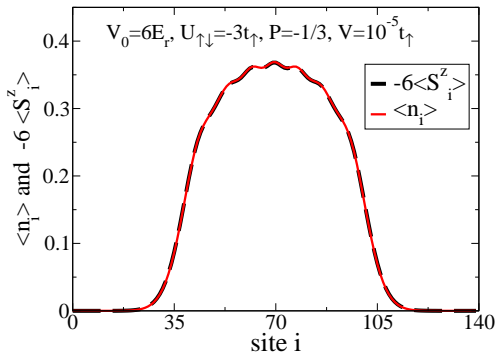}
\caption{Polarization and total density distribution in the $P=-1/3$ case for 
attractive $^{87}$Rb bosonic mixtures. Here, $N=21$.}
\label{f:bos_bal_dens}
\end{figure}

\begin{figure}[t]
\includegraphics[width=65mm]{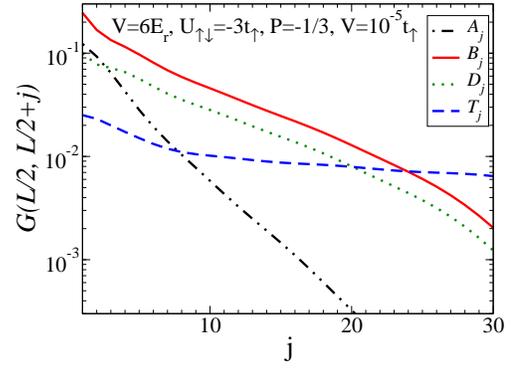}
\caption{Correlation functions in the middle of the trap in the $P=-1/3$ case
as a function of the distance with respect to the trap center: all correlations decay
exponentially except for $T$, pointing to a dominant trimer instability. Here, $A_j (B_j)$
denotes single particle correlation function for the $\uparrow (\downarrow)$ species.}
\label{f:bos_bal_corr}
\end{figure}

As we have seen in Sec.~\ref{sec:density}, fermionic gases with exactly 
commensurate polarization can stabilize a trimer phase over the entire
system, as shown from the density profiles in Fig.~\ref{fig:p_trim}. Since the trimer
instability is favored in the low-density regime, we focused on population
regimes where the typical total density in the middle of the trap fluctuates 
around $\langle n_i\rangle\lesssim0.5$. In Fig.~\ref{f:bos_bal_dens}, a typical density profile
is shown for $N=21$ particles in a $L=140$ site system: as in the fermionic
case, $\langle n_i\rangle $ and $-6\langle S^z_i\rangle$ coincide within numerical errors all over the 
system, thus providing a first signature of trimer formation. Another independent
evidence of trimer formation is then obtained by looking at the correlation function
decay in the central part of the system; in particular, in the trimer phase, both $B(x; L/2)$
and pairing correlations
\begin{equation}
D(x; L/2)=\langle b^{\dagger}_{\uparrow,L/2} b^{\dagger}_{\downarrow,L/2}b_{\uparrow,L/2+x}b_{\downarrow,L/2+x}\rangle
\end{equation}
are expected to decay exponential with $x$, whilst trimer correlations
\begin{eqnarray}
&&T(x; L/2)=\\
&&\langle b^{\dagger}_{\uparrow,L/2} b^{\dagger}_{\downarrow,L/2+1} b^{\dagger}_{\downarrow,L/2}b_{\uparrow,L/2+x}b_{\downarrow,L/2+x}b_{\downarrow,L/2+1+x}\rangle\nonumber
\end{eqnarray}
are expected to decay algebraically. In Fig.~\ref{f:bos_bal_corr}, the decay of correlation
functions in the middle of the system for the same parameter set of Fig.~\ref{f:bos_bal_dens}
clearly shows that, while both single particle and dimer correlations decay 
exponentially, the trimer correlations do not, further confirming the 
stability of the trimer phase in the inhomogeneous setup; moreover, $T(x;L/2)$
does not show any oscillating behavior with $x$, in accordance with the
bosonic nature of the three-body bound state~\footnote{Oscillation with 
respect to the effective Fermi momentum emerge in fermionic mixtures,
where the bound state has fermionic statistics: see Refs.~\onlinecite{roux11,
dalmonte2011}}.

\subsubsection{Incommensurate polarization}

As we have seen in Sec.~\ref{sec:density} for the case of fermions and in Sec.~\ref{s:trimer_onebcoherence} for hardcore bosons, 
even a very small deviation from commensurate
densities has a drastic effect on trimer phases.
 The consequences of $P\neq-1/3$ may be even more drastic in the 
bosonic case, which are not subject to the Pauli principle and can thus 
form higher density regions in the middle of the trap. 

\begin{figure}[t]
\includegraphics[width=65mm]{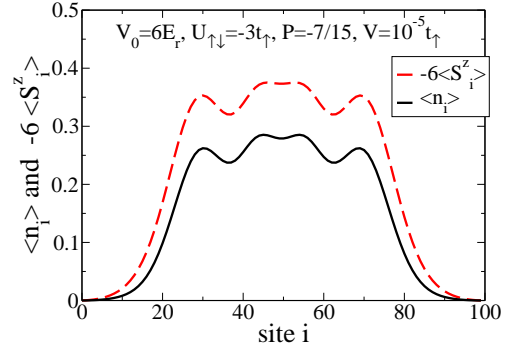}
\caption{Polarization and total density distribution in the $P=-7/15$ case for 
attractive $^{87}$Rb bosonic mixtures. The total number of particles in the system 
is $N=21$.}
\label{f:bos_unbal_dens}
\end{figure}

\begin{figure}[t]
\includegraphics[width=65mm]{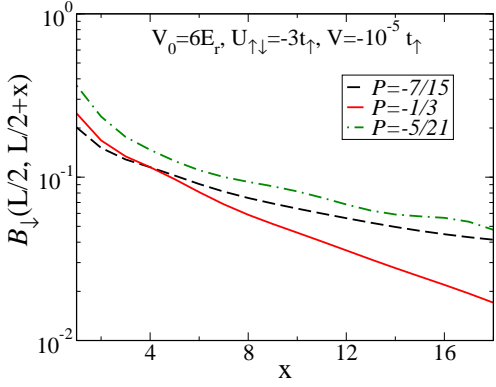}
\caption{Comparison between superfluid correlations of the heavy species at different
polarizations.}
\label{f:bos_unbal_corr}
\end{figure}

We have investigated the same parameter regime as in  Sec.~\ref{sec:fermions} 
in the presence of a minimal density imbalance, $\delta=P+1/3\simeq 0.1$, which
does also represent the typical experimental threshold of population control.
Similarly to the fermionic case, the trimer liquid is fragile with respect
to incommensurability. At first, one sees that the condition $2\langle n_{\uparrow,i}\rangle=
\langle n_{\downarrow,i}\rangle$ is not fulfilled in the middle of the trap: as shown
in Fig.~\ref{f:bos_unbal_dens},  there is a significant departure from commensurability
even for the smallest density imbalance considered. Secondly, correlation
functions are also strongly affected: in particular, single-particle correlations
change significantly from exponential to algebraic decay, signally the emergence
of quasi-condensation. In Fig.~\ref{f:bos_unbal_corr}, we compare the superfluid
correlation of the heavy species at different density imbalance for different
values of $P$: in both incommensurate cases, the decay in the center of the
trap is algebraic. 

\subsection{Momentum distribution signatures for soft-core bosons: finite-temperature effects}

\begin{figure}
\centering
\includegraphics[width=9cm]{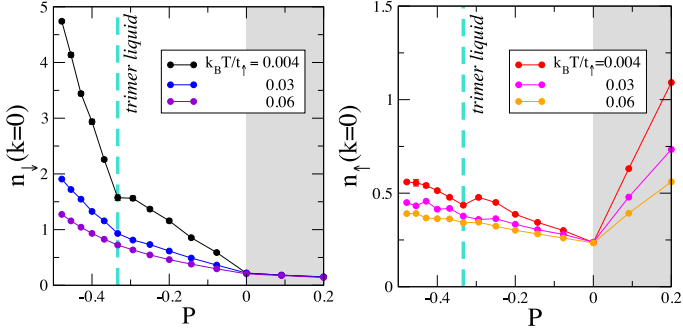}
\caption{\label{fig:trimerformation-nk0-T}(color online) Polarization dependence of the condensate peak at various temperatures for a trapped mixture of hardcore bosons with $N_{\uparrow}=12$ and increasing $N_{\downarrow}$. Other parameters are as in Fig.~\ref{fig:trimerformation-nk0}.}
\end{figure}

\begin{figure}
\centering
\includegraphics[width=9cm]{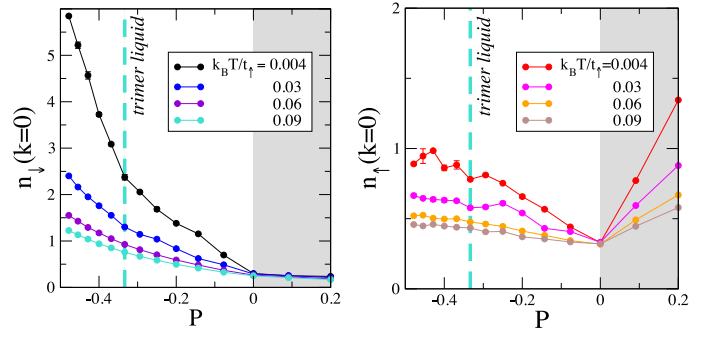}
\caption{\label{fig:trimerformation-nk0-softT}(color online) Polarization dependence of the condensate peak at various temperatures for a trapped mixture of softcore 
bosons with $N_{\uparrow}=12$ and increasing $N_{\downarrow}$. Here  $U_{\sigma\sigma}=10 t_\uparrow$ and $U=-3t_{\uparrow}$; all other parameters are as in 
Fig.~\ref{fig:trimerformation-nk0-T}. }
\end{figure}

Despite the fragility of trimer formation to slight departures from the $P=-1/3$ condition, a strong signature 
of this phenomenon is still observed in the evolution of the one-body coherence when $P$ is changed around
the value $-1/3$, as discussed in Sec.~\ref{s:trimer_onebcoherence}.
Here we discuss the robustness of this effect for finite temperatures, and for the softcore case. 
Figs.~\ref{fig:trimerformation-nk0-T} and \ref{fig:trimerformation-nk0-softT} show the height of the condensate peak with a varying
polarization $P$ for a mixture containing $N_{\uparrow}=12$ atoms and a variable number $N_{\downarrow}$
of $\downarrow$-atoms. Fig.~\ref{fig:trimerformation-nk0-T} refers to hardcore bosons at increasing temperatures, while 
Fig.~\ref{fig:trimerformation-nk0-softT} shows the case of softcore bosons with the same parameters as for $^{87}$Rb with
$V_0\approx 6 E_r$, and with $U = -3 t_{\uparrow}$. In both cases the trapping potential is very weak, 
$V_\uparrow = V_\downarrow =  1.8*10^{-5} ~t_{\uparrow}$. 

We observe that the distinct signatures of formation of dimer and trimer liquids -- the kinks in $n_{\sigma}(k=0)$ at $P=0$ and $P=-1/3$ 
respectively -- are still observed in the softcore case, albeit less clearly than in the hardcore case, particularly for what concerns the 
trimer kink. Both for the softcore and the hardcore case the kinks appear to be robust at fairly low temperatures, although the trimer kink is quickly rounded off as T reaches values in the range $0.05-0.1~ t_{\uparrow}/k_B$. This shows clearly that the observation of trimer formation with the current experimental diagnostics of time-of-flight measurements is indeed possible as far as the Hamiltonian parameters are concerned (\emph{e.g.} for $^{87}$Rb mixtures), but it requires extreme conditions of very low densities and very low temperatures. The two latter requirements are actually compatible, as very low temperatures can be in principle achieved by further evaporative cooling or algorithmic cooling \cite{popp06} of the atoms already loaded in the optical lattice.

\subsection{Summary: Homonuclear two-component Bose gases}

Binary bosonic mixtures represent a valuable setup where both dimer and trimer
liquid physics in presence of mass imbalance leaves strong signatures in the momentum distribution
accessible to experiments. In case of hard-core
intraspecies repulsion, equal density pairing in the central part of the trap is relatively stable
with respect to thermal effects, and may be observed in a wide range of polarizations
as long as the lighter species density is larger, that is $P>0$. Trimer physics is instead
more delicate: why in general soft-core interactions increase the trimer binding energy
at fixed mass imbalance, a very small departure from the commensurate condition $P=-1/3$
drastically changes both correlation functions and real space distribution of trapped systems. 
Nevertheless, sufficiently low temperatures and densities may indeed allow for the 
observation of such exotic liquids in a trapped gas by looking at the momentum distribution
of the heavier component in both hard-core and soft-core mixtures, a sharp signature being
the strong suppression of zero-momentum contribution as a function of $P$.

\section{Discussion and Summary}
\label{sec:summary}

Ultracold atomic gases represent ideal setups to explore unconventional superfluid states
of quantum low dimensional systems such as FFLO states and superfluids of composite particles. In this work, 
we have investigated binary bosonic and fermionic mixtures in the presence of both finite
spin- and mass-imbalance, taking into consideration typical experimental features such 
as the  inhomogeneity induced by trapping potentials and a finite temperature. In the first part,
we have shown how one can modify the effective mass imbalance by considering a proper
tuning of the underlying optical lattice potential, inducing different tunneling rates
for the different species. 
In the second part, taking full advantage of the detailed microscopic Hamiltonian
study of Sec.~\ref{sec:model}, we have carried out a combined DMRG and QMC numerical study
of both Fermi-Fermi and Bose-Bose mixtures for realistic experimental parameters. 
In the fermionic case, a finite mass-imbalance helps in stabilizing exactly paired phases
in presence of spin-imbalance: in sharp contrast with the equal mass case, a majority of 
light particles stabilizes an equal density region in the middle of the trap
in a broad regime of total densities and polarizations. This phase is also robust to
finite temperature effects, as corroborated by QMC simulations. 
The partially polarized phase is of the FFLO type, with a clear signature
in the momentum distribution of pairs. FFLO should be easier to see on the
$P<0$ side, where the heavy fermions are the majority species.
We can thus conclude that mass-imbalanced fermionic mixtures such as $^6$Li-$^{40}$K
represent a valuable setup to observe equal-density pairing and FFLO superfluidity in 1D 
systems under realistic experimental conditions. 
Large mass imbalance is known to lead to a richer showcase of superfluid
states known as composite liquids: as a case study, we have investigated the presence
and stability of trimer liquids in both bosonic and fermionic mass-imbalanced mixtures.
While the typical binding energy of trimer bound states is smaller than standard pairs,
 in the case of exactly commensurate densities trimer liquids are indeed
robust in inhomogeneous setups even in the presence of slightly different trapping potentials.
In the bosonic case, an experimental signature of such strongly correlated composite liquids
is provided by the single particle momentum distribution, which displays a sharp kink
in its polarization dependence even at finite (albeit small) temperature.
However, once one departs from $P=-1/3$, the trimer liquid is not stable, as signaled by both the
density distribution and the correlation function decay in the central part of the trap.
In summary, the experimental realization of trimer liquids in binary, mass-imbalanced mixtures 
represents a challenging task, as both low temperature and densities, combined with an
accurate control on spin-imbalance, will be necessary. Nevertheless, such phases may
indeed leave appreciable experimental signatures even on easily accessible observables
such as momentum distribution.

\acknowledgments
We thank R. Lutchyn and G. Orso for fruitful discussions.
F.H.-M., K.D. and U.S. acknowledge support form the Deutsche Forschungsgemeinschaft through FOR 801.
M.D. acknowledges support by the European Commission via the integrated project AQUTE. 
K.D. was supported by the National Research Foundation and the Ministry of Education, Singapore.
F.H.-M. and U. S. thank the KITP at UC Santa Barbara, where parts of this research were carried out, for its hospitality. This research was supported in part by the NSF under Grant. No. NSF PHY05-51164.
A.E.F. thanks NSF for support through Grant No. DMR-0955707.


\bibliographystyle{amsplain_nosort}

\end{document}